\documentclass[a4paper,11pt]{article}
\pdfoutput=1

\usepackage{jheppub}

\usepackage[T1]{fontenc}
\usepackage[mathletters]{ucs}
\usepackage{multicol}
\usepackage{multirow}
\usepackage{braket}

\title{Precision Higgs Width and Couplings \\ with a High Energy Muon Collider}

\preprint{YITP-SB-2023-22}

\author{Matthew Forslund}
\author{and Patrick Meade}
\affiliation{C. N. Yang Institute for Theoretical Physics,\\Stony Brook University, Stony Brook, NY 11794}

\emailAdd{matthew.forslund@stonybrook.edu, patrick.meade@stonybrook.edu}

\abstract{
The interpretation of Higgs data is typically based on different assumptions about whether there can be additional decay modes of the Higgs or if any couplings can be bounded by theoretical arguments. Going beyond these assumptions requires either a precision measurement of the Higgs width or an absolute measurement of a coupling to eliminate a flat direction in precision fits that occurs when $|g_{hVV}/g_{hVV}^{SM}|>1$, where $V=W^\pm, Z$.  In this paper we explore how well a high energy muon collider can test Higgs physics without having to make assumptions on the total width of the Higgs.  In particular, we investigate off-shell methods for Higgs production used at the LHC and searches for invisible decays of the Higgs to see how powerful they are at a muon collider.  We then investigate the theoretical requirements on a model which can exist in such a flat direction. Combining expected Higgs precision with other constraints, the most dangerous flat direction is described by generalized Georgi-Machacek models.  We find that by combining direct searches with Higgs precision, a high energy muon collider can robustly test single Higgs precision down to the $\mathcal{O}(.1\%)$ level without having to assume SM Higgs decays.  Furthermore, it allows one to bound new contributions to the width at the sub-percent level as well.  Finally, we comment on how even in this difficult flat direction for Higgs precision, a muon collider can robustly test or discover new physics in multiple ways.  Expanding beyond simple coupling modifiers/EFTs, there is a large region of parameter space that muon colliders can explore for EWSB that is not probed with only standard Higgs precision observables.
}

\begin{document} 
\maketitle

\section{Introduction}\label{sec:introduction}

A high energy muon collider is ideally suited to investigate the physics of Electroweak symmetry breaking (EWSB)~\cite{AlAli:2021let,Aime:2022flm,Black:2022cth,Buttazzo:2020uzc}, since ultimately both precision and energy are needed to explore its origins. Energy is needed to produce multi-Higgs boson processes that test the Higgs potential, increase the production cross section of single Higgs processes, to test the ``restored'' limit of EW symmetry and any source of any deviations from the standard model (SM) in the Higgs sector. Precision is needed to be able to test the couplings of the Higgs to other SM particles beyond the HL-LHC. While there exist strategies to investigate the physics of EWSB separately with an $e^+e^-$ precision factory~\cite{ILCInternationalDevelopmentTeam:2022izu,CEPCPhysicsStudyGroup:2022uwl,Bernardi:2022hny,Bai:2021rdg,Brunner:2022usy} followed by a high energy proton collider~\cite{Benedikt:2022kan}, a muon collider can provide both precision and energy in the same machine. Moreover, a muon collider at high energy is effectively an EW gauge boson collider~\cite{Buttazzo:2018qqp,Han:2020uid,Han:2021kes,AlAli:2021let} and thus is an ideal high energy machine for questions surrounding EWSB. 

A high energy muon collider has already been shown to have great potential for both single~\cite{Forslund:2022xjq,AlAli:2021let} and multi-Higgs measurements~\cite{Han:2020pif,Buttazzo:2020uzc,Chiesa:2020awd}. However, as with any collider study, one has to carefully treat how observables translate into actual knowledge of the underlying physics. In~\cite{Forslund:2022xjq}, a basic assumption was made that there are no additional decay channels for the SM Higgs boson. This allows one to interpret cross section measurements in either a $\kappa$-fit~\cite{LHCHiggsCrossSectionWorkingGroup:2012nn,LHCHiggsCrossSectionWorkingGroup:2013rie,deBlas:2019rxi} (specifically, ``$\kappa-0$'' with this assumption) or EFT fit in a self consistent manner without requiring an explicit Higgs width measurement, since any changes in the width are completely correlated with shifts in the couplings. Nevertheless, this may be too strong of an assumption, but then how well can you measure the properties of the Higgs without having to specify all possible BSM decay modes of the Higgs?  If we remain agnostic about new contributions to Higgs decays, then treating Higgs precision with coupling modifiers is still valid as long as the total width is also left as a free parameter. However, to then extract the precision on individual Higgs couplings requires additional information since any on-shell exclusive measurement is only sensitive to the combination 
\begin{equation}\label{eqn:parametric}
    \sigma(i\rightarrow H\rightarrow j)\sim \frac{g_i^2g_j^2}{\Gamma_H}.
\end{equation}
Therefore, extracting the couplings in full generality requires either an independent width measurement or an absolute measurement of one of the couplings. Without this, one can in principle confound precision measurements of couplings by hiding it in a flat direction where the couplings and the Higgs width are {\em increased} such that naively it looks like the SM, but there are actually large deviations to its properties~\cite{Azatov:2022kbs}.

Fortunately, there are both measurements that can be made and theoretical considerations which can be applied to understand whether the Higgs is SM-like and what its width is. For example, at the LHC, one can exploit gauge invariance of the SM to measure the effects of modified Higgs couplings from a highly off-shell Higgs contribution~\cite{Caola:2013yja, Campbell:2013una} to $VV$ scattering, where $V=W^\pm,Z$. This is independent of the Higgs width in the off-shell regime, and therefore can provide an absolute measurement of a coupling which removes the ambiguity. This has been carried out by ATLAS~\cite{ATLAS:2023dnm} and CMS~\cite{CMS:2022ley} thus far and there are projections that with the HL-LHC~\cite{Dawson:2022zbb} that claim a 17\% measurement uncertainty on the SM width can be achieved. While this is a remarkable achievement for the LHC, given that a direct width measurement is not remotely possible at the $\mathcal{O}(1)$ level\footnote{There is an additional LHC method exploiting interference in the $H\rightarrow \gamma \gamma$ on-shell rate~\cite{Campbell:2017rke} that likewise gives a subdominant precision.}, it ultimately sets a ceiling for how well you can interpret a measurement of Higgs couplings.

\begin{figure}[t]
    \centering
    \includegraphics[width=.495\textwidth]{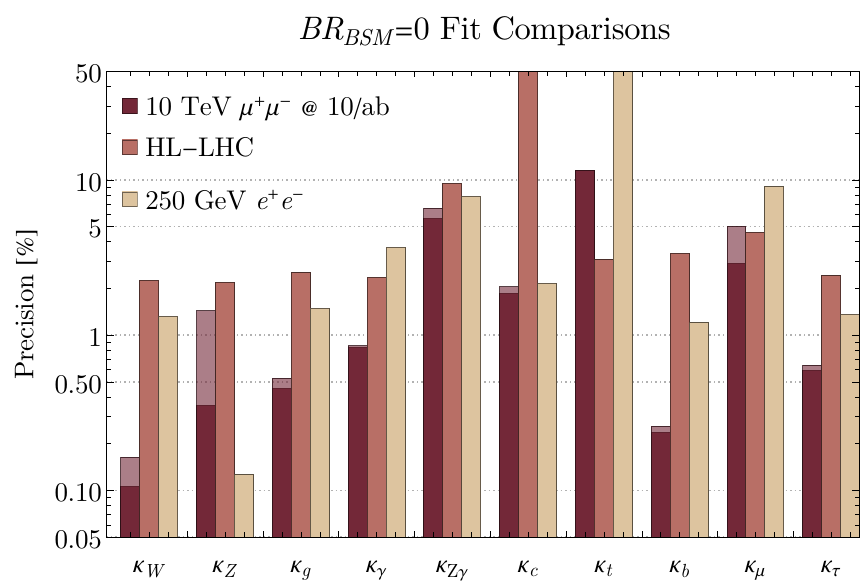}
    \includegraphics[width=.495\textwidth]{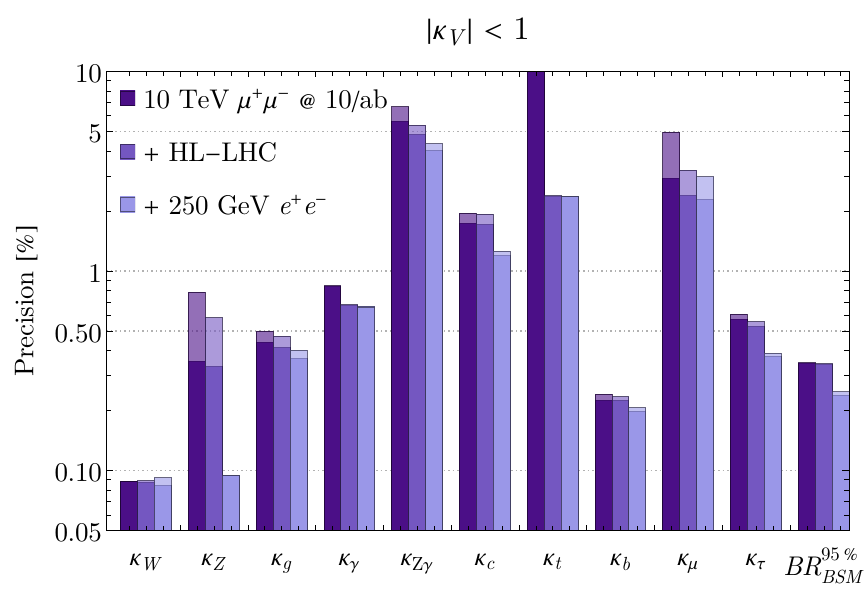}
    \caption{Fit results in the $\kappa$-framework using the on-shell results of~\cite{Forslund:2022xjq} with the assumptions to break the flat direction, where we use the fitting procedure described in Appendix~\ref{app:fits}. (left) A comparison of results for $BR_{inv}=0$ for a 10 TeV $\mu^+\mu^-$ collider, the HL-LHC, and a 250 GeV $e^+e^-$ collider. (right) Fit results with the assumption $|\kappa_V|<1$ for the muon collider alone, in combination with the HL-LHC, and in combination with a 250 GeV $e^+e^-$ collider. The transparent bars show the effect of removing forward tagging (see Appendix~\ref{app:methods}).}
    \label{fig:K0fits}
\end{figure}

The difficulty of having a ``width'' measurement with a substantially worse uncertainty than exclusive signal strengths is that a global fit will naturally have uncertainties on the couplings inherited from the width measurement. In particular, in the $\kappa$ framework one can treat all couplings as independent\footnote{Throughout this paper, we will consider the loop induced coupling modifiers $\kappa_g$, $\kappa_\gamma$, and $\kappa_{Z\gamma}$ as independent parameters to be fully agnostic to new states running in the loops. Specifying these in terms of the other $\kappa$'s would strictly increase precision. } and define the deviation from the standard model by a modifier $\kappa_i \equiv g_i/g_i^{SM}$ such that the on-shell signal strength of any given Higgs production and decay channel may be written
\begin{equation}
    \mu_{i\rightarrow H \rightarrow j} \equiv \frac{\sigma_{i\rightarrow H \rightarrow j}}{\sigma_{i\rightarrow H \rightarrow j}^{SM}} = \frac{\kappa^2_i\kappa^2_j}{\Gamma_{H}/\Gamma_H^{SM}}=\frac{\kappa^2_i\kappa^2_j}{\kappa_H^2}(1-BR_{BSM}), \qquad \kappa_H^2 \equiv \sum_i \frac{\kappa_i^2\Gamma_i}{\Gamma_H^{SM}},
\end{equation}
where $\sigma_{i\rightarrow H \rightarrow j }$ is the on-shell Higgs cross section in production channel $i$ and decay channel $j$, $BR_{BSM}$ is the sum of all BSM branching ratios of the Higgs, and $\Gamma_i$ is the partial width for the standard model decay $H\rightarrow i$. Here we have used the narrow-width approximation, $\Gamma_H \ll m_H$, which is justified by current LHC constraints on the total width~\cite{ATLAS:2023dnm,CMS:2022ley}. In this framework, if only exclusive signal strengths are measured, then the uncertainty on a given $\delta \kappa_i$ will naturally be limited by $\sim \delta \Gamma_H/4$. Therefore, for LHC results one often resorts to a $\kappa-0$ fit or adds an additional theory motivation. For example, the flat direction present in a global fit Eqn.~(\ref{eqn:parametric}) where the couplings and width are both increased can be explicitly seen if we assume a universal coupling modifier $\kappa_i=\kappa_H=\kappa$. In this case, the Higgs width scales as
\begin{equation}
    \frac{\Gamma_H}{\Gamma_H^{SM}} = \frac{\kappa^2}{1-BR_{BSM}},
\end{equation}
so that for any given channel, the on-shell signal strength becomes
\begin{equation}\label{eqn:flat}
    \mu_{i\rightarrow H \rightarrow j} = \frac{\sigma_{i\rightarrow H} \times BR_{H\rightarrow j}^{SM}}{\sigma_{i\rightarrow H}^{SM} \times BR_{H\rightarrow j}^{SM}} = \kappa^2(1-BR_{BSM}).
\end{equation}
For $\kappa > 1$, there is always a possible $BR_{BSM}$ to make all signal strengths $\mu_i = 1$, hence the flat direction in a fit. Clearly, if one assumes no BSM decay modes of the Higgs as in a $\kappa-0$ fit then this isn't an issue, {\em or} if one assumes that some of the $\kappa_i$ are bounded to be less than $1$. The latter is a commonly invoked by assuming any $|\kappa_V|\leq 1$, which may appear ad hoc but has theory motivations that we will discuss later. In Figure~\ref{fig:K0fits}, we show results for the $\kappa$ fit for these two assumptions for the 10 TeV $\mu^+\mu^-$ muon collider\footnote{We use total integrated luminosity benchmarks of 3 ab$^{-1}$ and 10 ab$^{-1}$ for the 3 TeV and 10 TeV $\mu^+\mu^-$ colliders, respectively.} and other representative colliders\footnote{For the 250 GeV $e^+e^-$ collider we have used CEPC inputs in~\cite{An:2018dwb}. Other 250 GeV options may give slightly different results depending on luminosities and run plans~\cite{Dawson:2022zbb}.}, both independently and in combination. \footnote{For details on the procedure used for all presented fits, see Appendix~\ref{app:fits}.}

\begin{figure}[t]
    \centering
    \includegraphics[width=\textwidth]{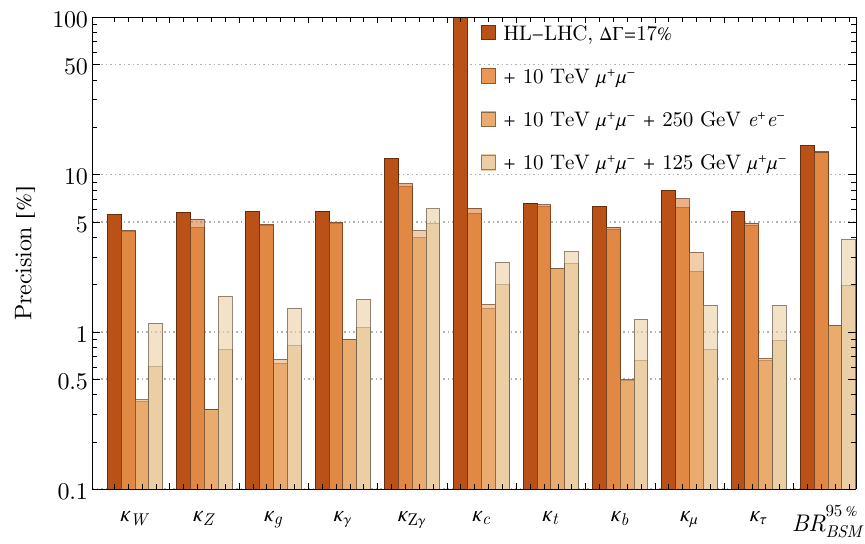}
    \caption{A demonstration of using the width measurement from another collider to resolve the flat direction in the fit, where we use the fitting procedure described in Appendix~\ref{app:fits}. We compare general fit results for the HL-LHC with its expected width precision of 17\%, the HL-LHC in combination with the on-shell 10 TeV $\mu^+\mu^-$ results, the on-shell 10 TeV $\mu^+\mu^-$ results combined with a 250 GeV $e^+e^-$ collider, and the on-shell 10 TeV $\mu^+\mu^-$ results combined with a 125 GeV $\mu^+\mu^-$ collider. Transparent bars show the combined differences from removing our 10 TeV forward muon tagging (see Appendix~\ref{app:methods}) and changing the luminosity of the 125 GeV collider between 20 fb$^{-1}$ and 5 fb$^{-1}$.}
    \label{fig:width17}
\end{figure}

A 10 TeV muon collider is clearly impressive and able to reach the $\mathcal{O}(.1\%)$ uncertainty independent of any other collider input if either of these assumptions hold, but if they don't, then the coupling measurement precision {\em could} be significantly degraded. To illustrate this, we show the result of the Higgs precision for a 10 TeV muon collider with additional BSM decay contributions assuming the ``width'' constraint comes from a {\em different} collider in Figure~\ref{fig:width17}. For example, one could use the HL-LHC projection just discussed and then, as is clearly seen, a high energy muon collider appears to be only marginally better than the HL-LHC, as expected based on our earlier comments. At an $e^+e^-$ Higgs factory, one can also make a precise ``absolute'' coupling measurement, by exploiting the fact that at $\sim~250$ GeV there is a dominant $ZH$ production mechanism that in combination with a ``clean'' environment allows for a high precision inclusive rather than exclusive cross section measurement. This can then translate into a roughly $\mathcal{O}(1\%)$ level measurement on the Higgs width, which is good enough to  approach the $\kappa-0$ precision if combined with a 10 TeV muon collider. Another possibility is for a direct width measurement from a threshold scan of the cross section that can in principle be performed at a 125 GeV muon collider, which also translates a roughly $\mathcal{O}(1\%)$   level width measurement~\cite{Barger:1995hr,Barger:1996jm,Han:2012rb,deBlas:2022aow}.

Figure~\ref{fig:width17} illustrates that a high energy muon collider, in combination with other future colliders can begin to re-approach the precision of a $\kappa-0$ or $\vert\kappa_V\vert<1$ fits in Figure~\ref{fig:K0fits}. However, it is still unclear whether a low energy Higgs factory would definitely occur before a high energy muon collider. Therefore, it is important to understand how precisely a high energy muon collider can test the Higgs independent of any additional inputs, and more importantly, if can it do better. To answer this we investigate a number of different routes, both standard methods applied to a muon collider as well as exploiting what can be learned directly using the significant direct energy reach of a 10 TeV muon collider.

\begin{figure}[t]
    \centering
    \includegraphics[width=\textwidth]{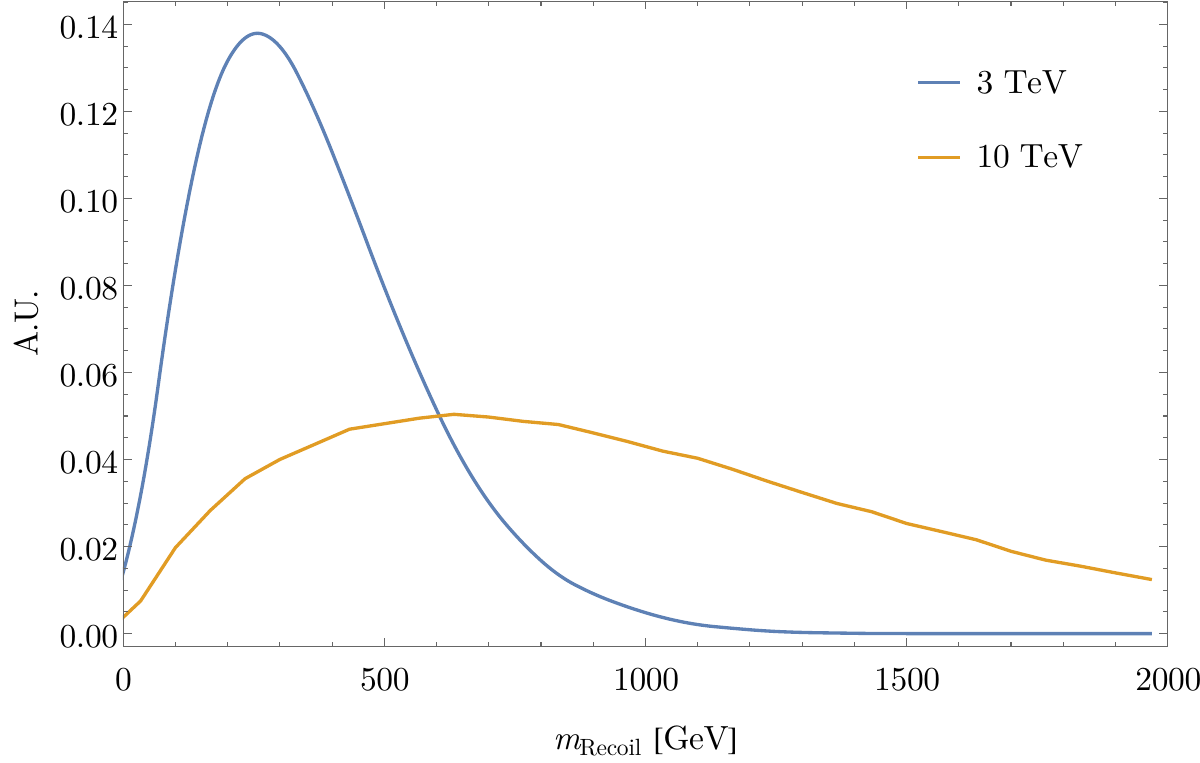}
    \caption{Normalised recoil mass distributions from the tagged forward muons in $\mu^+\mu^-\rightarrow \mu^+\mu^- H$ at 3 TeV and 10 TeV with a forward energy resolution of 10\%, using the methodology discussed in Appendix~\ref{app:methods} with an invisibly decaying $H$. Any peak at $m_H=125$ GeV is washed out by energy resolution effects, making an inclusive measurement extremely difficult, even at 3 TeV.}
    \label{fig:RecoilMass}
\end{figure}

At a high energy muon collider the production of Higgs bosons are dominated by vector boson fusion (VBF) production. Therefore, unlike a low energy $e^+e^-$ Higgs factory, the recoil mass method (see, for example, \cite{Bernardi:2022hny}) to obtain a precise inclusive Higgs measurement is quite difficult. The cross section for $ZH$ production is simply far too small at these energies to be useful, and performing a recoil mass measurement using $ZZ$-fusion Higgs production, $\mu^+\mu^- \rightarrow \mu^+\mu^- H$ requires an energy resolution on the forward muons far better than realistically attainable, as shown in Figure~\ref{fig:RecoilMass}.\footnote{For this comparison we have used the same event generation and detector assumptions as in the rest of our paper. See Appendix~\ref{app:methods} for details.} On the other hand, a muon collider is naturally suited to employ similar off-shell methods as the LHC. Off-shell methods have already been shown to enable a measurement on $y_t$ of 1.5\%~\cite{Chen:2022yiu,Liu:2023yrb}\footnote{See Appendix~\ref{app:fits} for details on how including this measurement changes our fit results.} at a 10 TeV muon collider, far better than attainable from on-shell $t\bar{t}H$ production~\cite{Forslund:2022xjq}. Applying these methods to $VV$ production to unambiguously fix $\kappa_V$ and remove the flat direction is a natural next step. In Section~\ref{sec:offshellanalysis} we outline in more detail how this method works and present our results. The off-shell method does require an assumption that the value of the coupling at the Higgs mass is the same as the value measured at high energies in $VV$ scattering. While this assumption is rather benign, it can still be tested directly at a high energy muon collider when one considers that the only loophole possible requires new physics coupled to the Higgs at low scales. Importantly, the only way to reduce the sensitivity shown in Figure~\ref{fig:K0fits} would be to have new physics that effectively exists along the flat direction of Eqn.~\ref{eqn:flat}. This requires {\em both} new BSM decay modes of the Higgs boson and a universal increase in single Higgs couplings, $|\kappa_i|>1$. 

Generating BSM decays of the Higgs is relatively straightforward through the Higgs portal; however, $|\kappa_V|>1$ is far more difficult to accomplish consistently. Given that the coupling precision of the $\kappa-0$ fit is at the level of $\mathcal{O}(.1-1\%)$, it would require a deviation of this order of magnitude for both $\kappa_V$ and new BSM Higgs branching fractions to obfuscate the existing Higgs precision results. To achieve a $|\kappa_V|>1$ at this level requires particular scalar states that mix with the Higgs at tree-level. For example, commonly studied singlet scalars or 2HDM models can be shown to strictly suppress $\kappa_V$~\cite{Logan:2014ppa}, which is why fits that assume $\vert\kappa_V\vert<1$ are theoretically natural. However, to ensure that the results for a muon collider are truly robust, we can go further and investigate the space of models that can generate $\vert\kappa_V\vert>1$, i.e. additional scalars coupled to the Higgs in representation of $SU(2)$ larger than the fundamental. This is a very narrow model building direction, because generically these representations violate the custodial $SU(2)_L \times SU(2)_R$ symmetry and cannot satisfy EW precision tests while also allowing for $\vert\kappa_V\vert>1$. The only model building direction that can accomplish this is the extension of so-called generalized George-Machacek models~\cite{GEORGI1985463,CHANOWITZ1985105,GALISON198426,PhysRevD.32.1780,Haber:1999zh,Chang:2012gn,Logan:2015xpa,Chiang:2018irv,Kundu:2021pcg} which incorporate multiple higher scalar $SU(2)$ representations with a potential that is custodially symmetric. Furthermore, after applying direct searches, models that are viable for $\vert\kappa_V\vert>1$ also require additional states for the BSM decay modes, creating a Rube Goldbergesque scenario to try to reduce the sensitivity of a 10 TeV muon collider. Nevertheless, one can investigate this direction thoroughly at a muon collider to test this hypothesis, and in the end, letting the Higgs width float arbitrarily can still be tested at a similar level of $\mathcal{O}(.1\%)$. In Section~\ref{sec:kv} we review the classes of models that can generate $\vert\kappa_V\vert>1$ and how they can be robustly tested. In Section~\ref{sec:BRinv} we give an example of the power of a high energy muon collider to test new decay modes of the Higgs with a specific example of invisible Higgs decays. This doesn't test all possible BSM Higgs decay modes, but for the flat direction to reduce the Higgs precision to the $\mathcal{O}(1\%)$ level it would require new decays modes accounting for $\mathcal{O}(10^5)$ Higgs decays at 10 TeV muon collider, which should be able to be discovered. We conclude in Section~\ref{sec:conclusion} with a review of how a 10 TeV muon collider on its own is a robust test of single Higgs precision down to the $\mathcal{O}(.1\%)$ level under the most general assumptions. This is achieved not solely through the standard Higgs fits, but by the fact that with a 10 TeV collider one can test BSM Higgs physics in multiple ways simultaneously. We include several appendices with some details omitted from the main text. In particular, in Appendix~\ref{app:methods} we discuss our event generation and detector assumptions used throughout the paper. Appendix~\ref{app:fits} describes our fitting procedures, and Appendix~\ref{Appendix:correlation} discusses the importance of correlations on our fits. Finally, in Appendix~\ref{app:fittabs} we include tables of the $\kappa$-fit results at 3 TeV and 10 TeV for all of the different assumptions discussed in the paper.

\begin{figure}[t]
    \centering
    \includegraphics[width=\textwidth]{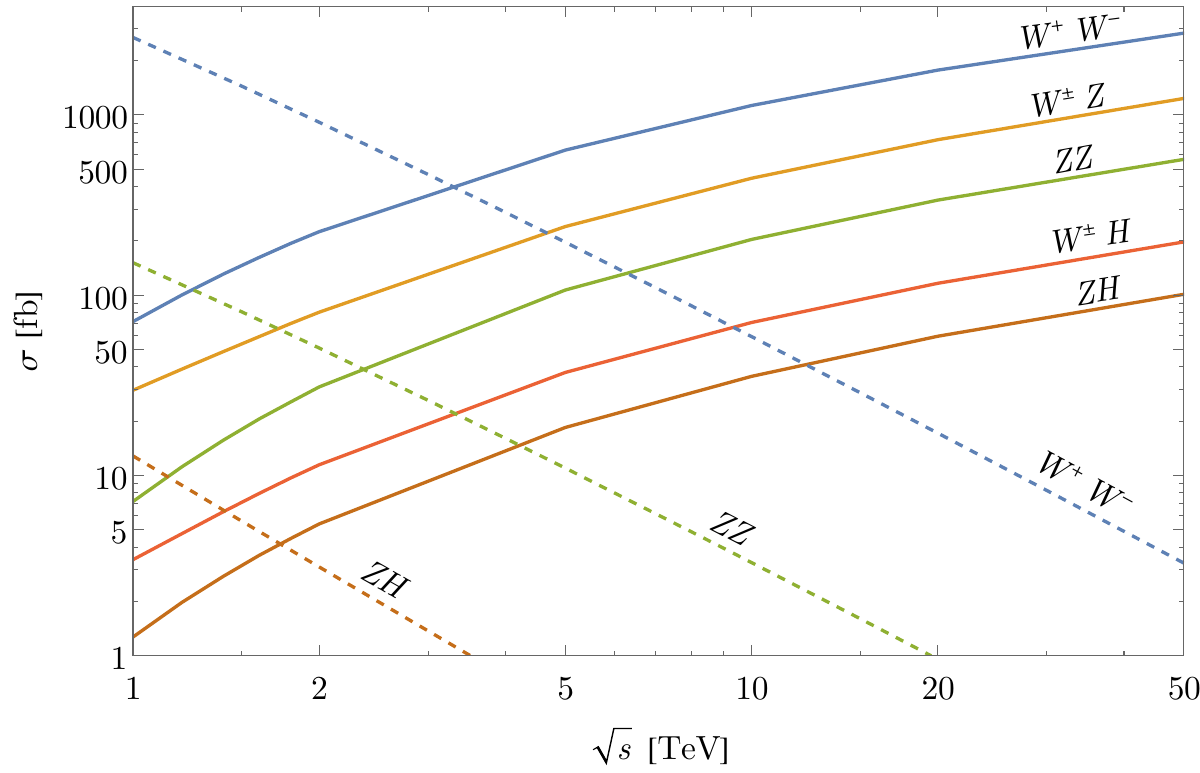}
    \caption{Cross sections as a function of center of mass energy for relevant diboson processes. Vector boson fusion processes, $(\mu^+\mu^-,\mu^\pm \nu_\mu, \bar{\nu}_\mu \nu_\mu)XX$ with $X=W^\pm , Z, H$, have solid lines, while the corresponding $s$-channel processes have dashed lines. Associated final state muons have a cut of $p_{T_\mu}>10$ GeV to regulate phase space singularities.}
    \label{fig:xsecs}
\end{figure}

\section{Off-shell analysis}\label{sec:offshellanalysis}

At a high energy muon collider, gauge boson scattering processes quickly become overwhelmingly dominant, making off-shell Higgs measurements much more promising than at the LHC. We show the cross sections computed with {\sc MadGraph5}~\cite{Alwall:2014} for the most important diboson processes as a function of CM energy in Figure~\ref{fig:xsecs}, where it is clear that the VBF cross sections are all quite large and are the dominant contributions to all relevant final states in most of phase space. This intuition fails at diboson invariant masses near our center of mass energy, where the $s$-channel processes with much smaller overall cross sections become dominant again and act as a cutoff to our $m_{VV}$ reach, as we will see. In the off-shell region, $\sqrt{\hat{s}}\gg m_H$ and the width drops out of the Higgs diagram contributions. Measuring it therefore resolves the degeneracy, since $\mu_{i\rightarrow H^* \rightarrow j}\equiv \sigma_{i\rightarrow H^* \rightarrow j}/\sigma_{i\rightarrow H^* \rightarrow j}^{SM} = \kappa_i^2\kappa_j^2$ so long as $\kappa_i(m_h)\kappa_j(m_h) = \kappa_i(\sqrt{\hat{s}})\kappa_j(\sqrt{\hat{s}})$, and no new BSM states contribute to the off-shell diagrams\footnote{These assumptions are related, since causing any meaningfully large change in the measured $\kappa_V$ as one goes to higher energies generally requires new states at least as light as those energies.}. Both of these are at least approximately true for a wide class of models, and any light states that would break this assumption would be well probed at a muon collider, as will be seen in Section~\ref{sec:kv}. Naively, the off-shell rate seems like it would be heavily suppressed and therefore difficult to measure to high precision. However, perturbative unitarity requires that $\kappa_V = 1$, as there is a delicate cancellation between the Higgs diagrams and the continuum that prevents the cross section from growing with energy. This is especially true for longitudinal electroweak gauge boson scattering where $\sigma_{V_LV_L\rightarrow V_LV_L}\propto \hat{s}^2$ if $\kappa_V \neq 1$. The $\hat{s}^2$ energy scaling of $VV$ scattering leads to $m_{VV}^4$ scaling of the $\mu^+\mu^-\rightarrow (\mu^+\mu^-,\mu^\pm\nu_\mu,\bar{\nu}\nu)VV$ differential cross section $d\sigma / dm_{VV}$ when varying $\kappa_V$, which allows measurements in the high $m_{VV}$ to be enhanced with respect to the naive intuition. 

We study the dominant decay channels of $4j$, $2\ell 2j$, and $\ell^\pm\nu_\ell jj$, since the low backgrounds at a lepton collider enable the hadronic channels to be used effectively. The comparatively low statistics of the fully leptonic decay modes make them unlikely to significantly increase the precision, so we do not consider them here. We note that while the attainable precision of the on-shell analysis~\cite{Forslund:2022xjq} in the hadronic channels was quite sensitive to the jet energy resolution, the same is not true here, as we are analysing a large continuum instead of separating resonances. Likewise, since we are looking at high energy final states, the beam-induced-backgrounds at muon colliders should not be relevant, as they give a diffuse low energy contribution. 

\begin{figure}[t]
    \centering
    \includegraphics[width=.495\textwidth]{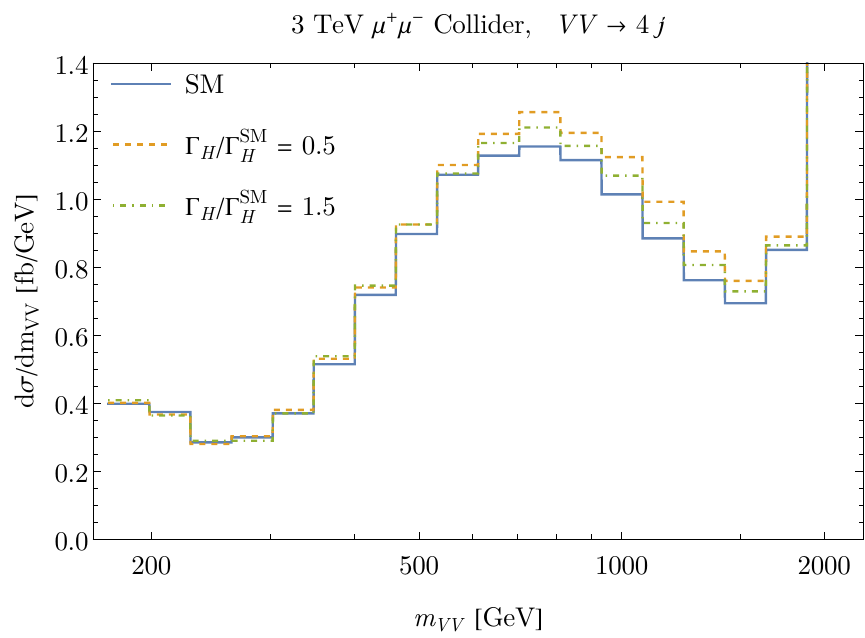}
    \includegraphics[width=.485\textwidth]{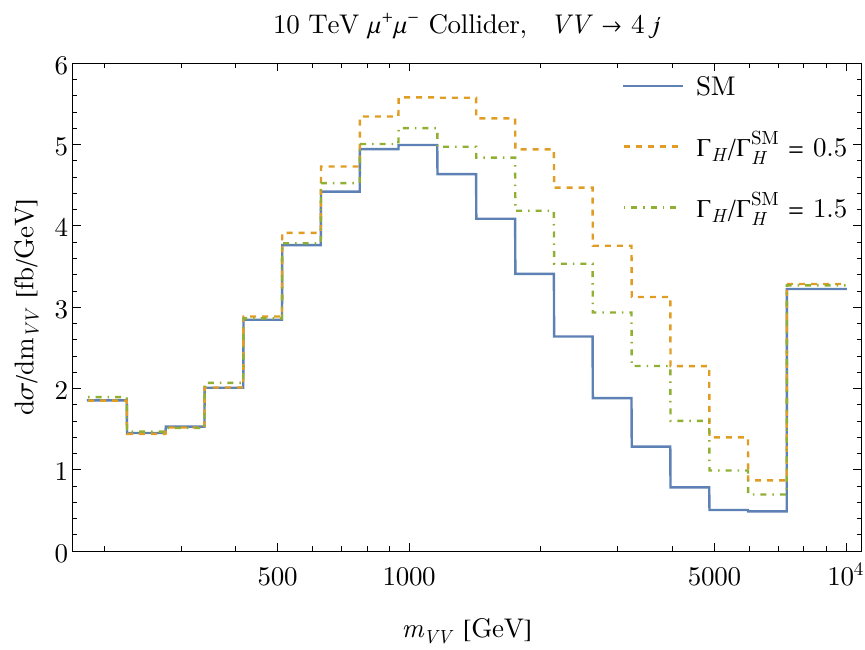}
    \caption{Total $d\sigma/dm_{VV}$ distributions for combined $VV\rightarrow VV\rightarrow 4j$ and relevant backgrounds at 3 TeV (left) and 10 TeV (right) after applying cuts. Changes in $\kappa_V$ that would correspond to a 50\% deviation in the width are shown over the standard model expectation. The lower $m_{VV}$ reach and higher $s$-channel backgrounds in the final bins at 3 TeV limit sensitivity compared to at 10 TeV.}
    \label{fig:m4j}
\end{figure}

We adopt a simple binned analysis, splitting the reconstructed $m_{VV}$ distribution for each channel into 20 bins\footnote{We have checked that these results are insensitive to the choice of number of bins.}, with smaller bin widths at lower invariant masses where the cross section is larger. For each process, we generate events with a wide range of $\kappa_W$ and $\kappa_Z$ and run all variations through showering and fast detector simulation (see Appendix~\ref{app:methods} for details). The number of events in every bin $k$ is then independently fit to a quadratic function of $\kappa_i \kappa_j$,
\begin{equation}\label{fit}
    N^k_{i\rightarrow j} = a^k\kappa_i^2\kappa_j^2  + b^k\kappa_i\kappa_j + c^k,
\end{equation}
where the large interference leads to a large $b^k$ coefficient for the high energy bins. The value of this function at $\kappa_i = \kappa_j = 1$ is taken to be our measured SM value for that bin. More sophisticated analyses can improve these results, but this serves as a reasonable starting point to match the on-shell results already presented. We consider the following backgrounds for each process, where $f$ is any of the SM fermions and again $X = W^\pm,Z,H$:
\begin{multicols}{2}
\begin{itemize}
    \item  $(\mu^+\mu^-,\mu^\pm \nu_\mu, \bar{\nu}_\mu \nu_\mu)f\bar{f}$
    \item  $(\mu^+\mu^-,\mu^\pm \nu_\mu, \bar{\nu}_\mu \nu_\mu)XX$
    \item  $(\mu^+\mu^-,\mu^\pm \nu_\mu, \bar{\nu}_\mu \nu_\mu)4j$ (QCD)
    \item  $t\bar{t}$
    \item  $XX$
    \item  $4j$ (QCD)
\end{itemize}
\end{multicols}
The processes with associated $(\mu^+\mu^-,\mu^\pm \nu_\mu, \bar{\nu}_\mu \nu_\mu)$ are VBF while the others are $s$-channel, which are important in different kinematic regions. Practically speaking, the only VBF $f\bar{f}$ processes remaining after invariant mass cuts are $t\bar{t}$ and $tb$. QCD backgrounds are highly subdominant at a muon collider compared to those from electroweak processes, though we include them in the $4j$ channel where they are relevant. Any additional processes such as off-shell $V$ induced $(\mu^+\mu^-,\mu^\pm \nu_\mu, \bar{\nu}_\mu \nu_\mu)4f$ are highly subdominant after our cuts and are neglected\footnote{This is in contrast to our on-shell study~\cite{Forslund:2022xjq}, where these processes were the dominant backgrounds for $\mu^+\mu^- \rightarrow (\mu^+\mu^-,\bar{\nu}_\mu \nu_\mu) H, \ H\rightarrow VV^*$ since our reconstructed $m_{VV}$ was below the $2m_V$ threshold.}. The overwhelming majority of the background comes from continuum $(\mu^+\mu^-,\mu^\pm \nu_\mu, \bar{\nu}_\mu \nu_\mu)V_TV_T$ and $s$-channel $V_T V_T$ processes, since our signal is longitudinally polarized with slightly different kinematic distributions. We impose some channel specific cuts to remove some of this continuum, although it remains the dominant background.

The majority of the statistics is in the $4j$ final state. We impose preselection cuts of $p_T>60$ GeV and $|\eta|<2.5$ to remove much of the backgrounds and minimize the effect of potentially neglected nearly collinear backgrounds. The jets are then paired together into two parents closest to the $Z$ mass. The two reconstructed parent bosons are required to satisfy $30<m_V^{min}<100$ GeV and $40<m_V^{max}<115$ GeV for the lighter and heavier reconstructed particle, respectively. These bounds are chosen to be rather loose, since as long as the lower bounds are sufficiently large to remove $jj$ backgrounds from photon induced processes, the dominant backgrounds are from continuum $V_TV_T$ with the same reconstructed dijet invariant masses. The reconstructed diboson $m_{VV}$ distribution is shown in Figure~\ref{fig:m4j} at both 3 and 10 TeV. The peaking at high $m_{VV}$ is a direct consequence of choosing such a strict $p_T$ cut. The enhancement due to the $\sigma \propto \hat{s}^2$ scaling when $\kappa_V \neq 1$ is clearly visible in the high $m_{VV}$ regions, especially at 10 TeV. The regime with $m_{VV}$ larger than shown in the plot is irrelevant due to the impact of $s$-channel backgrounds, especially $W^+W^-$, swamping out the signal at 3 TeV.

\begin{figure}[t]
    \centering
    \includegraphics[width=.495\textwidth]{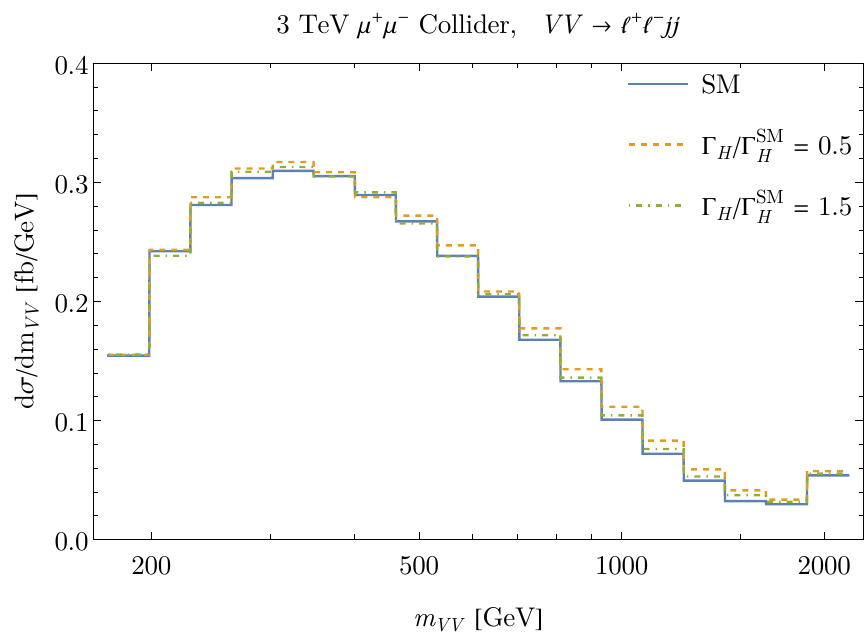}
    \includegraphics[width=.495\textwidth]{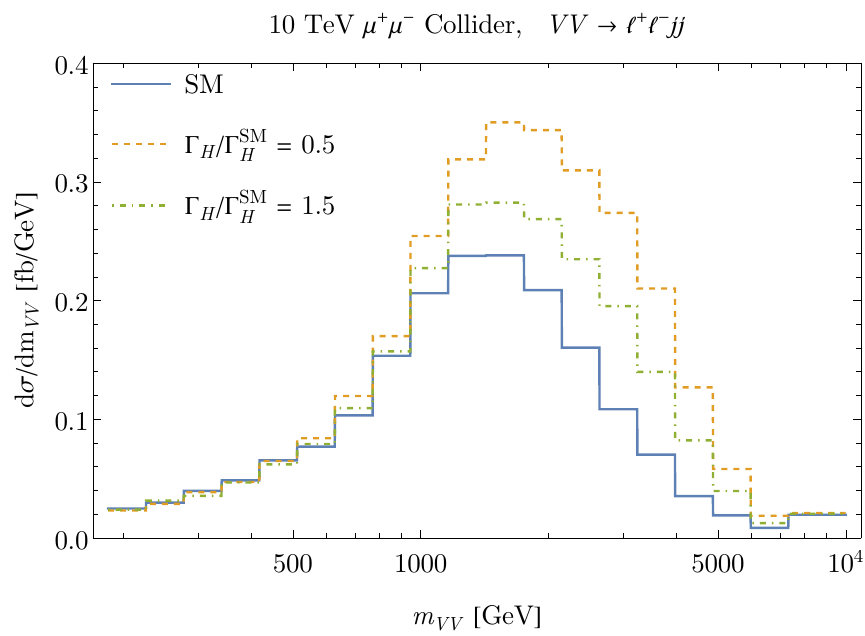}
    \caption{Total $d\sigma/dm_{VV}$ distributions for combined $VV\rightarrow VV\rightarrow 2l2j$ and relevant backgrounds at 3 TeV (left) and 10 TeV (right) after applying cuts. The angle cut at 10 TeV results in a peak at much higher $m_{VV}$. Changes in $\kappa_V$ that would correspond to a 50\% deviation in the width are shown over the standard model expectation.}
    \label{fig:m2l2j}
\end{figure}

For both the $2\ell2j$ and $\ell^\pm \nu_\ell jj$, we apply the preselection cut $|\eta|<2.5$ and a looser $p_T>20$ GeV cut since the presence of leptons reduces backgrounds. For $2\ell2j$, we apply invariant mass cuts of $70<m_{\ell\ell}<115$ GeV and $40<m_{jj}<115$ GeV for the parents reconstructed from the lepton and jet pairs, respectively. The lepton pair mass cut eliminates virtually all backgrounds without a $Z\rightarrow \ell^+ \ell^-$, making the only meaningful background contributions come from $VV\rightarrow ZV$ and $s$-channel $\mu^+\mu^-\rightarrow ZZ/ZH$. In particular, this means that the $s$-channel $W^+ W^-$ process does not contribute at all, while it was the dominant background for the $4j$ final state in the final bins. At 10 TeV, we also impose an angle cut on both pairs of $\theta_{\ell\ell},\theta_{jj} < 25^\circ$, while at 3 TeV such a similar cut does not improve precision. The $m_{VV}$ distributions at both 3 TeV and 10 TeV are shown in Figure~\ref{fig:m2l2j}, where the effects of the angle cut are immediately obvious, pushing the peak to much larger $m_{VV}$ values. For the $\ell^\pm \nu_\ell j j$ final state, the energy loss due to the neutrino from the $W$ decay makes it more challenging to reconstruct. We impose an invariant mass cut of $40<m_{jj}<115$ GeV, as well as cuts on the $p_T$ of the lepton and the dijet parent of $p_{T_\ell}<200(750)$ GeV and $p_{T_{jj}}<500(1200)$ GeV at 3(10) TeV, respectively. The exact values of these $p_T$ cuts are not particularly important, so long as they are sufficient to remove most of the $s$-channel backgrounds. We summarize the cuts for all channels in Table~\ref{tab:offshCuts}.

\begin{table}[h]
    \centering
    \renewcommand{\arraystretch}{1.3}
    \begin{tabular}{|c|c|c|c|}\hline
    Channel & $4j$ & $2\ell 2j$ & $\ell^\pm \nu_\ell j j$ \\ \hline 
    Cuts & \begin{tabular}{@{}c@{}}$|\eta_j| < 2.5$ \\ $p_{T_j} > 60$ \\ $30<m_V^{min}<100$ \\ $40<m_V^{max}<115$\end{tabular}  & \begin{tabular}{@{}c@{}}$|\eta_{\ell}|, |\eta_{j}| < 2.5$ \\ $p_{T_{\ell}} , p_{T_{j}} > 20$ \\ $70<m_{\ell\ell}<115$ \\ $40<m_{jj}<115$ \\ $\theta_{\ell\ell},\theta_{jj} < 25^\circ$ (10 TeV)\end{tabular} & \begin{tabular}{@{}c@{}}$|\eta_{\ell}|, |\eta_{j}| < 2.5$ \\ $p_{T_{\ell}} , p_{T_{j}} > 20$ \\ $40<m_{jj}<115$ \\ $p_{T_\ell}<200(750)$ \\ $p_{T_{jj}}<500(1200)$ \end{tabular} \\ \hline
    \end{tabular}
    \caption{A summary of the cuts applied to the different channels in our off-shell $VV$ analysis. Values in parenthesis were changed going from 3 TeV to 10 TeV. The angle cut for the $2\ell 2j$ final state is only at 10 TeV. All energies are in GeV.}
    \label{tab:offshCuts}
\end{table}

\begin{figure}[t]
    \centering
    \includegraphics[width=.495\textwidth]{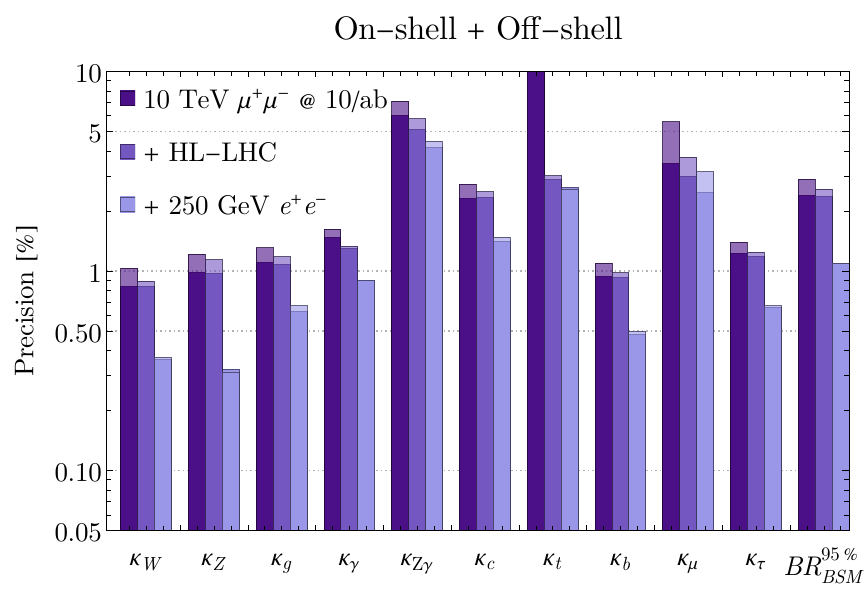}
    \includegraphics[width=.495\textwidth]{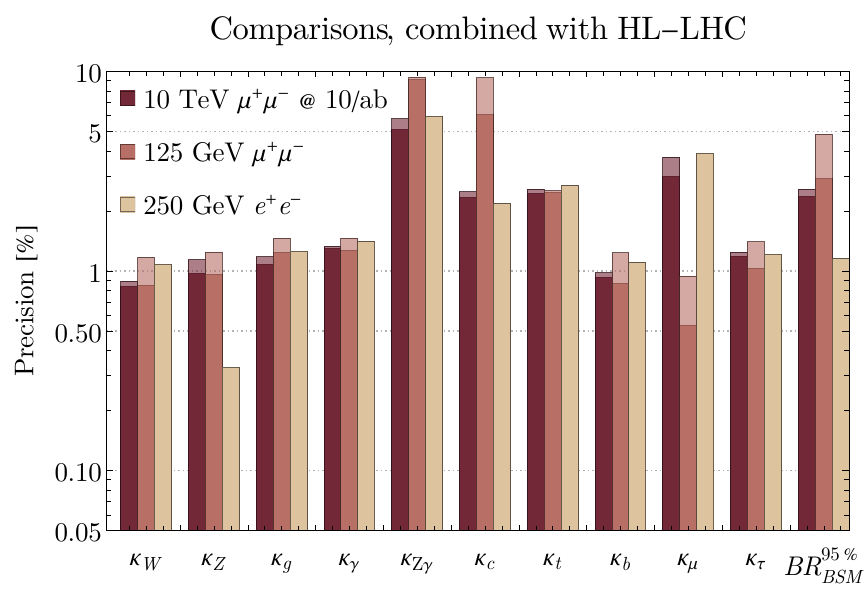}
    \caption{Fit results in the $\kappa$-framework including our off-shell observables at 10 TeV, where we use the fitting procedure described in Appendix~\ref{app:fits}. We present results (left) alone and in combination with the HL-LHC and a 250 GeV $e^+e^-$ collider, or (right) compared to a 125 GeV $\mu^+\mu^-$ collider and a 250 GeV $e^+e^-$ collider, all combined with the HL-LHC. The transparent bars show the effect of removing forward tagging for the 10 TeV $\mu^+\mu^-$ collider (see Appendix~\ref{app:methods}) and the effect of a reduction of luminosity from 20 fb$^{-1}$ to 5 fb$^{-1}$ for the 125 GeV $\mu^+\mu^-$ collider.}
    \label{fig:KOffshFits}
\end{figure}

We then input the bins for each final state as individual observables in HEPfit, in a similar manner to how the on-shell inputs are included (see Appendix~\ref{app:fits} for details). This allows us to do a fully general $\kappa$ fit, without the assumptions necessary before. We show results for this fit in Figure~\ref{fig:KOffshFits} for the 10 TeV muon collider alone and in combination with the HL-LHC and a 250 GeV $e^+e^-$ collider, as well as comparisons with a 250 GeV $e^+e^-$ collider and the 125 GeV $\mu^+\mu^-$ collider. It is worth noting that due to the inherently asymmetrical nature of our off-shell constraints, as well as the fact that $BR_{BSM}\geq 0$, our resulting posteriors are not Gaussians centred at the SM predictions, and there is a strong correlation between $BR_{BSM}$ and $\kappa_W$ (see Appendix~\ref{Appendix:correlation}), an artifact of the flat direction. All of the precisions we present for these off-shell fits are therefore the upper 68\% confidence band of each parameter's marginalised posterior distribution. For the muon collider alone, these fits yield a width precision of 3.4\% at 10 TeV and 24\% at 3 TeV. The 3 TeV numbers are not competitive with the HL-LHC, and we will therefore not discuss them much more in the text, although we include them in our tabulated fit results in Appendix~\ref{app:fittabs}.

\section{What can generate \texorpdfstring{$|\kappa_V| > 1$}{kV > 1}?}\label{sec:kv}

From the fully general off-shell $\kappa$ fit, we obtain a precision on $\kappa_W$ of $0.84\%$, substantially worse than our $0.1\%$ number when assuming $|\kappa_V| < 1$. This worse number is \textit{only} relevant when working with a model where $|\kappa_V| > 1$.  Therefore it is a natural question to ask, what space of QFT can populate this region?  Once this space is delineated, we can then ask the question, are we limited to the off-shell results or are there sufficient constraints such that we recover the precision of the $\kappa-0$ or $|\kappa_V| < 1$ fits?  The reason why this question is particularly important for a muon collider is the high energy reach.  For a ``standard'' 250 GeV Higgs factory, as long as the states are slightly above the EW scale they can be integrated out and the EFT or $\kappa$ prescription effectively tells the full story.  However, with at 10 TeV muon collider, treating the off-shell measurement of $\kappa_V$ and the on-shell measurement together is fraught with difficulty unless the new physics that causes any deviation in the Higgs sector is sufficiently heavy.  With a high energy muon collider that provides both precision and energy, one has to be careful in understanding the parameter space to determine its true precision and to not be limited by the formalism of lower energy precision experiments. We therefore want to ask, in the space of realizable QFTs where $|\kappa_V| > 1$, after the direct search bounds are taken into account so that $\kappa$ or EFT fits are self consistent, are they still limited by the off-shell precision?

From a model building perspective, to our knowledge, there are only two ways to accomplish this.  The first method is by introducing new $SU(2)_L$ scalar multiplets that contribute to electroweak symmetry breaking. These multiplets must be larger than doublets or they cannot generate $|\kappa_V|>1$~\cite{Logan:2014ppa}.  The second is if a composite Higgs model (CHM) is based on a non-compact symmetry group~\cite{Alonso:2016btr}.  In this case
\begin{equation}\label{eqn:ncchm}
\kappa_V=\sqrt{1+v^2/f^2},  
\end{equation}
where $f$ is the symmetry breaking scale which could naively be bounded to the multi-TeV scale with Higgs precision alone at a muon collider.  However, as pointed out in~\cite{Liu:2016idz}, while a non-compact CHM can be a consistent EFT, it cannot be UV completed by a unitary QFT.  Furthermore, it would require adding new decay modes to survive in the flat direction and a UV completion to properly asses the reach of a muon collider.  It is therefore not clear whether this is a viable QFT to interpret Higgs results, so we instead focus on large $SU(2)_L$ multiplets in this section. 

Before considering Higgs precision, electroweak precision constraints must first be satisfied. In particular, the ratio of $W$ and $Z$ boson masses, $\rho \equiv m_W^2 / (m_Z^2 \cos^2{\theta_W})$, is constrained to be very near to one~\cite{Workman:2022ynf}; in other words, the custodial $SU(2)_L \times SU(2)_R$ symmetry must be preserved at tree-level by the addition of new scalar multiplets. For an extended theory with multiple complex scalars, each with weak isospin $J_i$, hypercharge $Y_i$, and vacuum expectation value $v_i$, the total contribution to the $\rho$ parameter at tree-level is given by~\cite{Gunion:1989we}
\begin{equation}\label{eq:rho}
    \rho = \frac{\sum_i \left[J_i (J_i + 1)-Y_i^2\right]v_i^2}{\sum_i 2Y_i^2 v_i^2} .
\end{equation}
Any additional scalars that do not give $\rho=1$ must have extremely small vacuum expectation values to remain viable, and therefore cannot meaningfully contribute to Higgs precision. A scalar singlet with $Y=0$ or a scalar doublet with $Y=1/2$ both yield $\rho=1$, but cannot give $|\kappa_V| > 1$. After doublets, the next single multiplet solution preserving $\rho=1$ is a $Y=2$ scalar septet. However, adding such a septet that obtains a nonzero vacuum expectation value breaks an accidental global $U(1)$ symmetry.  This generates a massless Goldstone boson coupling to fermions, which is clearly ruled out. Removing this Goldstone boson is needed to make the model phenomenologically   viable~\cite{Hisano:2013sn,Kanemura:2013mc,Alvarado:2014jva}, for instance by adding a higher dimension operator to break the symmetry or gauging the $U(1)$. We will discuss this option further, but it turns out that avoiding the Goldstone renders Higgs precision back to the 
$\mathcal{O}(.1\%)$ level. Any other single multiplet solutions to $\rho=1$ violate perturbative unitarity due to their large weak charges~\cite{Hally:2012pu} and will therefore not be considered here.

The only other possibility for larger scalar $SU(2)_L$ representations is to add multiple scalars with a custodial symmetry preserving potential so that their contributions to \eqref{eq:rho} combine to give $\rho=1$. This is known as a Georgi-Machacek (GM) model, and while it was first pointed out for triplets~\cite{GEORGI1985463,CHANOWITZ1985105}, it can be straightforwardly extended to higher multiplets~\cite{GALISON198426,PhysRevD.32.1780,Haber:1999zh,Chang:2012gn,Logan:2015xpa,Chiang:2018irv,Kundu:2021pcg} as well. These avoid the Goldstone boson problem and have rich phenomenology, although hypercharge explicitly breaks the custodial symmetry~\cite{Gunion:1990dt,Blasi:2017xmc,Keeshan:2018ypw}, necessitating a UV completion appearing anywhere from a few TeV to $\mathcal{O}(100)$ TeV depending on model parameters to satisfy electroweak precision constraints. Any other method of adding large scalar multiplets while preserving $\rho=1$ would require extreme fine tuning. Large multiplets necessarily have a plethora of new states to search for, including singly and doubly charged scalars that can be effectively searched for at a high energy muon collider.

\subsection{A minimal example: the Georgi-Machacek model}\label{sec:GM}

To demonstrate the power of a muon collider in testing theories where $|\kappa_V| > 1$, we will start with an example and consider the simplest GM model before discussing more general implications. The GM model has been explored extensively in the literature over the last several decades~\cite{PhysRevD.42.1673,PhysRevD.43.2322,Gunion:1989we,Aoki:2007ah,Godfrey:2010qb,Logan:2010en,Low:2010jp,Low:2012rj,Chiang:2012cn,Falkowski:2012vh,Killick:2013mya,Englert:2013zpa,Englert:2013wga,Hartling:2014zca,Chiang:2014hia,Chiang:2014bia,Godunov:2014waa,Hartling:2014aga,Hartling:2014xma,Chiang:2015kka,Chiang:2015rva,Chiang:2015amq,Degrande:2017naf,Das:2018vkv,Ghosh:2019qie,Ismail:2020zoz,Wang:2022okq,Bairi:2022adc,deLima:2022yvn,Chakraborti:2023mya}. We will follow the conventions in~\cite{Hartling:2014xma} in what follows. The scalar field content of the GM model consists of the usual standard model Higgs doublet $(\phi^+,\phi^0)$, with an additional real triplet $(\xi^+,\xi^0,\xi^-)$ and complex triplet $(\chi^{++},\chi^+,\chi)$ with hypercharge $Y=0$ and $Y=1$, respectively. The fields may be written as a bi-doublet and a bi-triplet under $SU(2)_L \times SU(2)_R$ as
\begin{equation}
    \Phi = \begin{pmatrix} \phi^{0*} & \phi^+ \\ -\phi^{+*} & \phi^0\end{pmatrix}, \qquad \qquad 
    X = \begin{pmatrix} \chi^{0*} & \xi^+ & \chi^{++} \\ -\chi^{+*} & \xi^0 & \chi^+ \\ \chi^{++*} & -\xi^{+*} & \chi^0 \end{pmatrix}.
\end{equation}
The vacuum expectation values (vevs) for the two scalar multiplets are given by $\braket{\phi^0} = v_\phi$ and $\braket{\chi^0} = \braket{\xi^0} = v_\chi$, where custodial symmetry enforces $\braket{\chi^0} = \braket{\xi^0}$. The scalar kinetic terms are
\begin{equation}
   \mathcal{L} \supset \textrm{Tr}[(D^\mu \Phi)^\dagger D_\mu \Phi] + \textrm{Tr}[(D^\mu X)^\dagger D_\mu X]
\end{equation}
with the covariant derivatives defined in the usual way as
\begin{align}
\begin{split}
    D^\mu\Phi &= \partial^\mu + i g W_a^\mu \tau_a \Phi + i g' B^\mu \Phi, \\
    D^\mu X &= \partial^\mu + i g W_a^\mu t_a X + i g' B^\mu t_3 X, \\
\end{split}
\end{align}
where $\tau_a = \sigma_a /2$ as usual, and the $3\times 3$ generators $t_a$ are given by
\begin{equation}
    t_1 = \frac{1}{\sqrt{2}}\begin{pmatrix} 0 & 1 & 0 \\ 1 & 0 & 1 \\ 0 & 1 & 0\end{pmatrix}, \qquad t_2 = \frac{1}{\sqrt{2}}\begin{pmatrix} 0 & -i & 0 \\ i & 0 & -i \\ 0 & i & 0\end{pmatrix}, \qquad t_3 = \frac{1}{\sqrt{2}}\begin{pmatrix} 1 & 0 & 0 \\ 0 & 0 & 0 \\ 0 & 0 & -1\end{pmatrix}.
\end{equation}
After $X$ and $\Phi$ obtain vevs, electroweak symmetry breaking proceeds as usual, with the total vev fixed by measurements to be $(\sqrt{2} G_F)^{-1} = v^2 = v_\phi^2 + 8 v_\chi^2$, which lets us define 
\begin{equation}
    c_H \equiv \cos \theta_H = \frac{v_\phi}{v}, \qquad s_H \equiv \sin \theta_H = \frac{\sqrt{8}v_\chi}{v}.
\end{equation}
 The most general custodially symmetric scalar potential is given by
\begin{align}
    \begin{split}
        V(\Phi, X) =& \frac{\mu_2^2}{2} \textrm{Tr}(\Phi^\dagger \Phi) + \frac{\mu_3^2}{2} \textrm{Tr}(X^\dagger X) + \lambda_1 \textrm{Tr}[(\Phi^\dagger \Phi)]^2 + \lambda_2 \textrm{Tr}(\Phi^\dagger \Phi) \textrm{Tr}(X^\dagger X) \\
        & + \lambda_3\textrm{Tr}(X^\dagger X X^\dagger X) + \lambda_4 \textrm{Tr}[(X^\dagger X)]^2 - \lambda_5\textrm{Tr}(\Phi^\dagger \tau_a \Phi \tau_b)\textrm{Tr}( X^\dagger t_a X t_b) \\
        & - M_1 \textrm{Tr}(\Phi^\dagger \tau_a \Phi \tau_b) (UXU^\dagger)_{ab} - M_2 \textrm{Tr}(X^\dagger t_a X t_b) (UXU^\dagger)_{ab}
    \end{split}
\end{align}
where the last two terms in particular are necessary to make the model compatible with current LHC constraints~\cite{deLima:2022yvn}. The matrix $U$ rotates $X$ into the Cartesian basis and is given by 
\begin{equation}
    U = \begin{pmatrix}
        -\frac{1}{\sqrt{2}} & 0 & \frac{1}{\sqrt{2}} \\ -\frac{i}{\sqrt{2}} & 0 & -\frac{i}{\sqrt{2}} \\ 0 & 1 & 0
    \end{pmatrix}.
\end{equation}
After EWSB, in the gauge basis, there is a custodial fiveplet, a triplet, and two singlets defined by
\begin{align}
    \begin{split}
        &H_5^{++} = \chi^{++}, \qquad H_5^+ = \frac{1}{\sqrt{2}}(\chi^+ - \xi^+), \qquad H_5^0 = \sqrt{\frac{2}{3}}\xi^0 - \sqrt{\frac{1}{3}}\chi^{0,r}, \\ &H_3^+ = -s_H \phi^+ + c_H \frac{1}{\sqrt{2}}(\chi^+ + \xi^+), \qquad H_3^0 = -s_H \phi^{0,i} + c_H \chi^{0,i} \\
        &H_1^0 = \phi^{0,r} \\
        &H_1^{0\prime} = \sqrt{\frac{1}{3}} \xi^0 +\sqrt{\frac{2}{3}} \chi^{0,r}.
    \end{split}
\end{align}
where the superscripts $i$ and $r$ refer to the real and imaginary parts of the relevant neutral fields. Note that since the fiveplet does not contain any of the $SU(2)_L$ doublet $\phi$, it does not couple to fermions. In the mass basis, the singlets mix to become
\begin{equation}
    h = c_\alpha H_0^1- s_\alpha H_0^{1\prime},\qquad \qquad H = s_\alpha H_0^{1} + c_\alpha H_0^{1\prime},
\end{equation}
with $c_\alpha \equiv \cos{\alpha}$, $s_\alpha \equiv \sin{\alpha}$, and one of $h$ or $H$ the observed $125$ GeV Higgs. The modification of the $g_{hVV}$ coupling, the parameter we are primarily interested in, is given by 
\begin{align}
\begin{split}
    \kappa_{hVV} =& c_{\alpha} c_{H} - \sqrt{8/3} s_{\alpha} s_{H}, \\
    \kappa_{HVV} =& s_{\alpha} c_{H} + \sqrt{8/3} c_{\alpha} s_{H},
\end{split}
\end{align}
where the modification is the same for both $\kappa_Z$ and $\kappa_W$ at tree-level. Since the scalar triplets cannot couple to the fermions through any renormalizable interaction, the Yukawa sector is the same as the SM in the gauge basis. In the mass basis, one finds the coupling modifiers
\begin{equation}
    \kappa_{hff} =  \frac{c_\alpha v}{v_\phi} = \frac{c_\alpha}{c_H}, \qquad  \kappa_{Hff} = \frac{s_\alpha v}{v_\phi} =  \frac{s_\alpha}{c_H}.
\end{equation}
As we approach decoupling, $\mu_3^2 \gg \mu_2^2$, we may integrate out the heavy triplets. Only the trilinear interaction $M_1$ contributes to $\kappa_V$ and $\kappa_f$ at tree-level, since it is the only term linear in a heavy field. We may rewrite this term as
\begin{equation}
    \mathcal{L} \supset M_1 H^\dagger \tau^a H \xi^a +( M_1/\sqrt{2} )(H^T (i \sigma_2) \tau^a H \chi^a + \textrm{h.c.}),
\end{equation}
where $H$ is the SM Higgs doublet, $\chi$ is the complex triplet, and $\xi$ is the real triplet, all written as vectors. Integrating out the real scalar $\xi$ and complex scalar $\chi$ yields at tree-level~\cite{Corbett:2017ieo,Anisha:2021hgc}
\begin{align}
\begin{split}
    \mathcal{L}^\xi_{eff} &\supset - \frac{M_1^2}{2\mu_3^4}\mathcal{O}_{HD} + \frac{M_1^2}{8\mu_3^4} \mathcal{O}_{H} + \frac{M_1^2}{2\mu_3^4}\mathcal{O}_{R}\\
    \mathcal{L}^\chi_{eff} &\supset \frac{M_1^2}{2\mu_3^4}\mathcal{O}_{HD} + \frac{M_1^2}{2\mu_3^4}\mathcal{O}_{R}\\
\end{split}
\end{align}
where we have only written the dimension 6 operators modifying $\kappa_V$ and $\kappa_f$, and we use the notation
\begin{equation}
    \mathcal{O}_R = |H|^2 |D_\mu H|^2, \qquad \mathcal{O}_{HD} = |H^\dagger D_\mu H|^2, \qquad \mathcal{O}_{H} = (\partial_\mu |H|^2)^2.
\end{equation}
The $\mathcal{O}_{HD}$ terms must cancel as a result of custodial symmetry. After electroweak symmetry breaking, the remaining two operators yield terms proportional to $(v^2/4)(\partial_\mu h)^2$, giving
\begin{equation}\label{eq:kvDec}
    \kappa_V^{dec} \approx 1+\frac{3}{8} \frac{M_1^2 v^2}{\mu_3^4}, \qquad \kappa_f^{dec} \approx 1-\frac{1}{8} \frac{M_1^2 v^2}{\mu_3^4},
\end{equation}
which matches the result computed in the full model~\cite{Hartling:2014zca}. Importantly, as we approaches decoupling, $|\kappa_f| < 1$ while $|\kappa_V| > 1$, so even with a $BR_{BSM}$, there is no flat direction. The maximum allowed size of these coupling deviations can be found from perturbative unitarity of the quartic couplings, translated into a bound on $M_1/\mu_3$. To see this, note that in deriving the mass eigenstates of the GM model using $m_h$ as an input, one can eliminate $\lambda_1$ in terms of $m_h$. In the decoupling limit, this relation is given by~\cite{Hartling:2014zca} 
\begin{equation}
    \lambda_1 \approx \frac{m_h^2}{8v^2} + \frac{3M_1^2}{32 \mu_3^2},
\end{equation}
which can likewise be obtained in the EFT from the coefficient of the $|H^\dagger H|^2$ term. In the UV, perturbative unitarity of the full scalar scattering matrix at high energies yields $\lambda_1\leq \pi/3$, which translates to an upper bound $M_1/\mu_3 \lesssim 3.3$.

There are a number of existing constraints on the GM model from current LHC data which are conveniently included in GMCalc~\cite{Hartling:2014xma}. While most available parameter space exists for $m_5 \gtrsim 400$ GeV, some points survive with masses below 200 GeV, an unfortunate result of the existing LHC constraint on $pp\rightarrow H_5^{++}H_5^{--} \rightarrow 4W$ stopping at masses of $200$ GeV~\cite{Ismail:2020kqz,ATLAS:2021jol}. A future extension of these analyses to lower masses would likely rule out this mass window. That being said, a dedicated analysis may not even be necessary, with the luminosity of the HL-LHC. The cross section for $pp\rightarrow H_5^{++}H_5^{--}$ becomes very large as $m_5$ becomes small, and the final state of interest is quite unique. Even at low masses, each $H_5^{\pm \pm}$ predominantly decays to two off-shell $W^\pm$ bosons, resulting in an abundance of events such as $pp\rightarrow e^+e^+\mu^-\mu^-+inv$, which are very clean even at the LHC. Any excess of these events would appear in the validation regions of the SUSY search analysis in~\cite{ATLAS:2021yyr} as an excess. As a rough estimate of the resulting constraints from the SUSY search, we take the expected uncertainty in VR0 as present statistical uncertainty and scale it by the future HL-LHC luminosity. Using leading order NNPDF2.3 pdfs~\cite{Ball:2013hta}, we generate events for $pp\rightarrow H_5^{++}H_5^{--}\rightarrow 4\ell 4\nu$ at leading order for a variety of masses and run them through the ATLAS detector fast sim card included with {\sc Delphes} after showering. We impose the same set of cuts to the output as in~\cite{ATLAS:2021yyr} and use the resulting cross sections and efficiencies to obtain the resulting constraints in GMCalc. We find that even this simple non-dedicated search would eliminate nearly all surviving data points with $m_5 < 200$ GeV.

We are now in a position to implement direct searches at the muon collider itself. We consider two search channels\footnote{$W^+W^-$ fusion $H_5^0\rightarrow \gamma \gamma$ was similarly checked, however it is never the dominant constraint over any $m_5$ region once the HL-LHC SUSY search is included since the decay is only important for the low $m_5$ region. We have also checked the Higgstrahlung process $\mu^+\mu^- \rightarrow H_5^0 Z, \  Z\rightarrow \ell^+\ell^-$ with recoil mass cuts; however, it is significantly weaker than the combination of VBF $H_5^{0,\pm}$ production and perturbative unitarity, even at high masses. Including hadronic $Z$ decays and the additional process $\mu^+\mu^- \rightarrow W^\pm H_5^\mp$ would improve this constraint somewhat, but it is beyond our scope.} at the 10 TeV muon collider, again using the methodology described in Appendix~\ref{app:methods}: $\mu^+\mu^-\rightarrow (\mu^+\mu^-,\mu^\pm \nu_\mu, \bar{\nu}_\mu \nu_\mu)H_5^{0,\pm},$ $H_5^{0,\pm} \rightarrow ZZ/ZW^\pm \rightarrow \ell^+\ell^- j j$, and $H_5^{\pm\pm}$ pair production, $\mu^+\mu^- \rightarrow H_5^{++}H_5^{--}\rightarrow 3\ell 3\nu 2j$. The latter process is a clean signal and is produced with no $s_H$ suppression factor, so it yields the dominant constraint over the vast majority of parameter space. The former VBF production modes come along with a factor $s_H^2$ but have a higher mass reach due to only one heavy scalar needing to be produced.

For $H_5^{\pm\pm}$ pair production, we do a very simple analysis where we require all $p_{T_{\ell,j}}>80$ GeV to remove VBF backgrounds, $40<m_{jj}<200$ GeV to be consistent with a $W^\pm$ decay, $2000<m_{3\ell 2j}<9000$ GeV, and remove any events with a same flavor $\ell^+\ell^-$ pair with mass $m_{\ell\ell} < 110$ GeV to suppress $Z$ decays. We do not do any binning, and instead take the $2\sigma$ upper limit to be the statistical limit from the SM backgrounds passing these loose cuts. Clearly, more optimisation could do a much better job here, but even the simplest unbinned cut-and-count limits removes the overwhelming majority of currently allowed parameter space. A more sophisticated multi-channel analysis can likely push this constraint close to the 5 TeV kinematic limit.

For the VBF $\ell^+\ell^- j j$ search, we require $p_T>20$ GeV and $|\eta|<2.5$ for both leptons and jets, and bin in increments roughly the size of the reconstructed resonance, between 60-200 GeV, broader at higher energies. We impose additional cuts of $5<\min(m_{\ell\ell},m_{jj})<100$ GeV, $15<\max(m_{\ell\ell},m_{jj})<100$ GeV when the $Z$ bosons are off-shell, and tighten the cuts to the same as in Section~\ref{sec:offshellanalysis} once past threshold. We do not try to optimise the binning or cuts further, as any more optimised analysis will depend on detector and beam effects not included in our fast sim. The limit is taken to be purely the $2\sigma$ statistical limit from the SM backgrounds for each bin.

\begin{figure}[t]
    \centering
    \includegraphics[width=\textwidth]{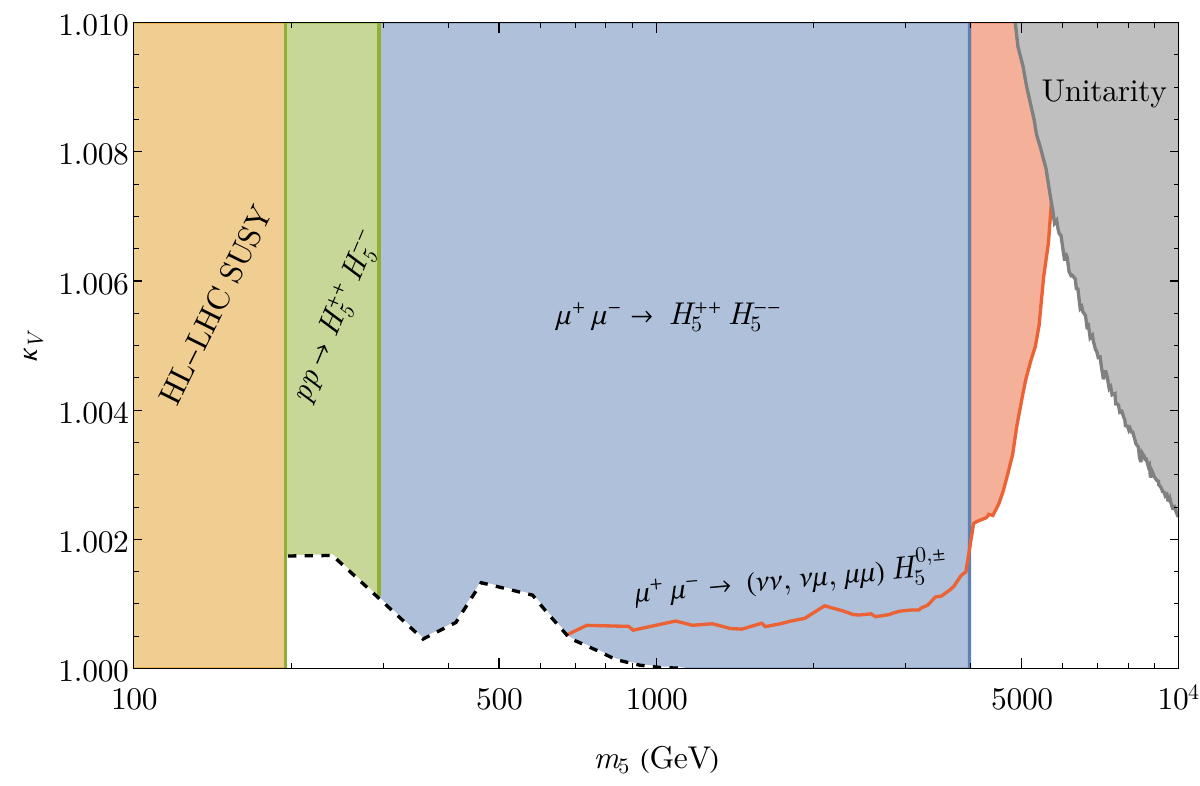}
    \caption{Expected direct search constraints from the HL-LHC and a 10 TeV $\mu^+\mu^-$ collider on the $|\kappa_V|>1$ regime of the Georgi-Machacek model. The orange region is the estimated HL-LHC reach from SUSY searches in multilepton final states, as discussed in the text. The green region is the current LHC constraint on $H^{++}H^{--}$ pair production~\cite{ATLAS:2021jol}. The blue and red regions are our estimated constraints from $H^{++}H^{--}$ pair production and VBF single $H^0$ and $H^\pm$ production at a 10 TeV $\mu^+\mu^-$ collider, respectively. The grey region shows the bound from perturbative unitarity of the quartic couplings. The allowed dashed region at $200\lesssim m_5 \lesssim 1000$ GeV is from very rare points where $H_5 \rightarrow H_3 H_3$ or $H_5 \rightarrow V H_3$ decays are dominant, and would likely be constrained by additional searches for $H_3 \rightarrow f\bar{f}$ final states. The maximum allowed $\kappa_V$ after imposing these searches is 1.007 at $m_5 \approx 6$ TeV.}
    \label{fig:GMconst}
\end{figure}

The results after implementing these constraints in GMCalc are shown in Figure~\ref{fig:GMconst}. The orange shows the previously mentioned SUSY search constraint scaled up to luminosity of the HL-LHC. The green band is excluded by the current LHC $pp\rightarrow H_5^{++}H_5^{--}$ constraints~\cite{ATLAS:2021jol}. The HL-LHC will push this constraint further right, up to roughly 600 GeV. The blue shows our 10 TeV $H_5^{\pm\pm}$ pair production constraint, which extends up to masses of about 4 TeV. In red-orange, our VBF constraints are shown, which extend a bit further than the pair production limit\footnote{Note that the step in this limit at $\sim$4 TeV is a physical consequence of increasing the bin size to 200 GeV, and not a statistical artifact.}. The gray region shows the unitarity bound on $\lambda_1$. The remaining white regions are allowed, where the small window at low masses is from very rare data points where the $H_5^{\pm \pm}$'s dominantly decay to other scalars. These points will be put under tension as the current LHC $pp\rightarrow H_5^{++}H_5^{--}$ constraints are improved with more data, and the region will likely shrink substantially by the end of the HL-LHC\footnote{This window at low masses can also be constrained in generalised GM models with larger multiplets by using Drell-Yan data at the LHC~\cite{Rainwater:2007qa,Alves:2014cda,Gross:2016ioi,DiLuzio:2018jwd} or a future muon collider~\cite{DiLuzio:2018jwd}.}. These additional scalars decay predominantly via either $H^\pm_3\rightarrow W^\pm Z$ or $H_3^\pm\rightarrow tb$, making for distinctive final states that are even easier to see than those we have considered. Dedicated searches would therefore almost certainly completely rule out this window at a muon collider. A number of further channels for direct searches could improve all of these constraints at a muon collider, such as $ZZ$ fusion processes and searches for the custodial triplet states. A comprehensive direct search program in all relevant final states is beyond the scope of this paper, but even our first order analysis presented here shows the qualitative features we are interested in. In particular, for masses below our off-shell binning, direct searches are far more constraining than the off-shell $\kappa$ limits, and force us to live in the decoupling limit. Since the decoupling limit implies $|\kappa_f|<1$, the fit with this assumption in Figure~\ref{fig:GMfits} applies directly to the remaining allowed high-mass region of the GM model.

\subsection{Universal implications}

\begin{figure}[t]
    \centering
    \includegraphics[width=\textwidth]{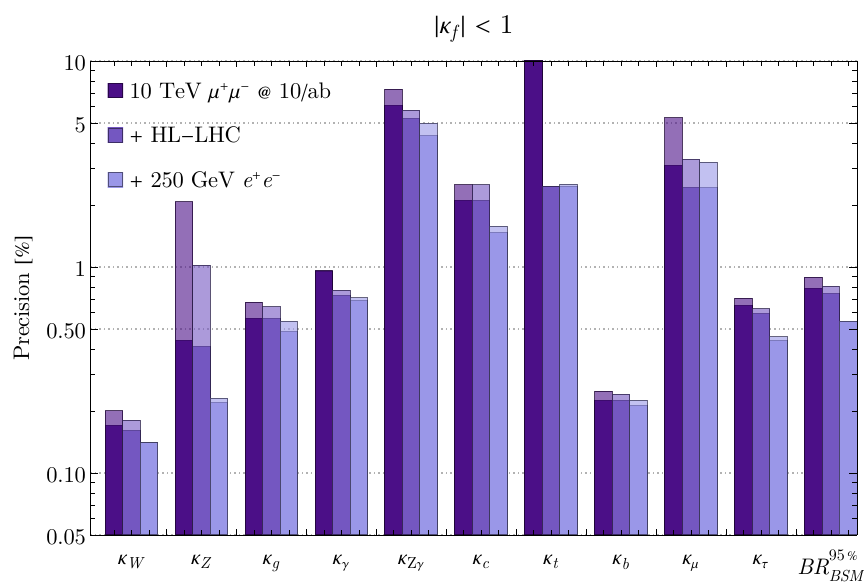}
    \caption{Fit results using a prior $|\kappa_f| < 1$ allowed by the decoupling limit of GM models, as discussed in the text. We use the fitting procedure described in Appendix~\ref{app:fits}. Results are shown for the muon collider alone, in combination with the HL-LHC, and in combination with the HL-LHC and a 250 GeV $e^+e^-$ collider. Transparent bars show the effect of removing forward tagging (see Appendix~\ref{app:methods}).}
    \label{fig:GMfits}
\end{figure}

Now that we have considered the constraints on the Georgi-Machacek model, let us see what can be learned about its generalisations. There are only three generalised GM models that are allowed by perturbative unitarity of transverse $SU(2)_L$ gauge boson scattering~\cite{Logan:2015xpa}: the custodial quartet~\cite{Durieux:2022hbu}, the quintet, and the hextet. All of these have a custodial fiveplet state after EWSB and mass diagonalisation, which can be constrained in the exact same way as described above. In fact, all of the direct search bounds in the $\kappa_V$ plane are identical for any of these models, since $H_5^{++}H_5^{--}$ production is independent of all model parameters, and while VBF constraints in the $s_H$ plane change, in the $\kappa_V$ plane they do not~\cite{Logan:2015xpa}. Direct searches therefore send any generalised GM model into the decoupling limit at a high energy muon collider. However, of these models, only the custodial quartet has a decoupling limit since it is the only one that can have an interaction with the Higgs linear in the heavy field. The quintet and hextet either would not be able to contribute to electroweak symmetry breaking or would be completely ruled out at a 10 TeV muon collider, just like the $\mathbb{Z}_2$--symmetric GM model~\cite{deLima:2022yvn}, and so we do not need to consider them further. 

The custodial quartet consists of a hypercharge $1/2$ quartet $S_{1/2}$ and hypercharge $3/2$ quartet $S_{3/2}$ coupling to the standard model doublet with terms $\mathcal{L} \supset \frac{\lambda_S}{\sqrt{3}} H_i H_j H_k S^{ijk *}_{3/2} + \lambda_S H_i H_j \widetilde{H}_k S^{ijk *}_{1/2}$, where we have written both as symmetric three-index representations of $SU(2)_L$. Since the coupling is quartic, the leading contributions to $\kappa_V$ and $\kappa_f$ appear at dimension 6 at one-loop order, and at dimension 8 at tree-level. They are given by~\cite{Durieux:2022hbu}
\begin{align}\label{eq:kvQuad}
\begin{split}
    \kappa_V &= 1+\frac{\lambda_S^2 v^2}{M^2}\left(\frac{1}{4\pi^2}+\frac{4 v^2}{3 M^2}\right), \\
    \kappa_f &= 1 - \frac{\lambda_S^2 v^2}{M^2}\left(\frac{1}{12\pi^2} + \frac{v^2}{3M^2} \right).
\end{split}
\end{align}
Notice that once again, $|\kappa_f| < 1$, and so the fit with this prior in Figure~\ref{fig:GMfits} to break the flat direction gives the appropriate $\kappa$ bounds. Likewise, perturbative unitarity of a quartic coupling $\lambda_S$ provides an upper bound to these coefficients, analogous to the unitarity bound on $M_1/\mu_3$ in the GM model. Explicitly computing this bound from the full scattering matrix, however, is unnecessary. In contrast to the triplet, for the quadruplet, while $\kappa_V$ and $\kappa_f$ are suppressed by a loop factor, the deviation in the trilinear Higgs self coupling is not, and is instead generated at tree-level dimension 6~\cite{Durieux:2022hbu}:
\begin{equation}
    \kappa_3 \equiv \frac{g_{hhh}}{g_{hhh}^{SM}} = 1-\frac{4}{3} \frac{\lambda_S^2v^4}{ M^2 m_h^2} .
\end{equation}
This means that for any large $\kappa_V$, there will be a hugely enhanced $\kappa_3$, which will be constrained to the 5\% level at a 10 TeV muon collider~\cite{Buttazzo:2020uzc,Han:2020pif}. At energies above our 4 TeV $H_5^{++}H_5^{--}$ bound where the maximal $\kappa_V \lesssim 1.007$ was found for the GM model, the custodial quartet would be constrained to $\kappa_V \lesssim 1.0003$ (or $0.03\%$) from this self-coupling constraint, using the above expressions for $\kappa_V$ and $\kappa_3$. The custodial quartet would therefore exclusively be \emph{more} constrained than the GM model. The constraints on the quartet in the $(\kappa_V, m_5)$ plane would look identical to our Figure~\ref{fig:GMconst} other than a differing unitarity bound and the bound from $\kappa_3$ cutting off $\kappa_V \lesssim 1.0003$. These two models are the full set of generalised Georgi-Machacek models generating $|\kappa_V|>1$ that need be considered, and both satisfy $|\kappa_f|<1$ after direct searches, allowing this fit assumption to break the flat direction. 

One may wonder about new electroweak states that do not contribute to EWSB and have no couplings linear in the heavy field, yet cause a deviation in $\kappa_V$. A scalar multiplet may couple to the standard model Higgs via an interaction\footnote{This neglects the interaction $(\chi^\dagger \chi) (H^\dagger H)$ and interactions with the gauge fields since they cannot generate $|\kappa_V| > 1$. If the multiplet has hypercharge 1/2, one can additionally write $(\widetilde{\chi}^\dagger T_\chi^a \chi)(H^\dagger T_H^a \widetilde{H})$ which we ignore.} $\mathcal{L}\supset - \lambda (\chi^\dagger T_\chi^a \chi)(H^\dagger T_H^a H)$, which generically leads to a $\kappa_W > 1$. If we integrate out such a multiplet with weak isospin $J$ and mass $M$, we find relevant terms
\begin{equation}
    \mathcal{L}\supset \frac{1}{768\pi^2 M^2}\left[\frac{\lambda^2 }{3}J(J+1)(2J+1)(4\mathcal{O}_{R}-4\mathcal{O}_{HD}+\mathcal{O}_{H})\right].
\end{equation}
These contributions are highly suppressed, as we may have guessed. The contributions to the Higgs couplings can then be computed as in~\cite{Durieux:2022hbu}. After considering direct searches, which will strongly constrain any electroweak charged states~\cite{Han:2022ubw}, even saturating perturbative unitarity will not result in a deviation in $\kappa_V$ of more than 1.007. For a concrete example, consider a septet, $n=7$. Saturating the perturbative unitarity bound $\lambda \leq 6.11$~\cite{Earl:2013jsa}, one finds $\delta \kappa^{n=7}_W \lesssim 0.57 v^2/M^2$ and $\delta \kappa^{n=7}_Z = 0$. A deviation of at least $\kappa_W = 1.007$ would require $M\lesssim 2.2$ TeV, which would be ruled out by direct searches at a muon collider~\cite{Han:2022ubw}\footnote{The specific direct searches for such a scenario would be somewhat different than these results, since such a large $\lambda$ would generate a larger mass splitting, leading to more energetic decay products which are easier to observe. Note that the lifetimes would be much shorter, so the disappearing track searches would not apply. Additional production mechanisms would also open up as a result of this interaction, providing more search channels. Nonetheless,~\cite{Han:2022ubw} (without disappearing tracks) provides a rough lower bound for what to expect for the reach.}. Note also that there is no flat direction in this scenario: $\kappa_Z \leq 1$ and $\kappa_f \leq 1$ independently of any model parameters. This asymmetry between $\kappa_W$ and $\kappa_Z$ also manifests as a contribution to the oblique $T$ parameter of $\alpha T = c_{HD} v^2/(2M^2)$, which can immediately be translated into a bound $\kappa_W \lesssim 1.002$ for $ T \lesssim 0.1$~\cite{Workman:2022ynf}.

For the $Y=2$ scalar septet, while a full analysis is beyond our scope, we may still draw some conclusions. The renormalisable couplings are captured by the above loop discussion, so we only need to consider the new effects when the septet gets a vev. As we have mentioned already, when this happens, an accidental $U(1)$ symmetry is broken yielding a massless Goldstone boson which must be removed either by using a higher dimensional operator to induce the vev or by gauging the accidental $U(1)$ symmetry. The septet vev will allow for the decays $H^{\pm\pm}\rightarrow W^{\pm}W^\pm$ and make our GM direct search bound apply, $M\gtrsim 4$ TeV, further enhanced by pair production of the higher charged scalars. The higher dimensional operator that is usually used~\cite{Hisano:2013sn,Alvarado:2014jva}, $\chi H^5 H^* $, gives the septet a vev $v_7 \sim v^6/(\Lambda^3 M^2)$. This lets us estimate the maximal $\kappa_V$ after our direct searches. In the most conservative scenario, $\Lambda \gtrsim M \gtrsim$ 4 TeV, and $|\kappa_V-1| \approx 4(v_7/v)\lesssim 10^{-6}$, several orders of magnitude smaller than our $\kappa_V$ fit sensitivities. In the case where the accidental $U(1)$ is gauged (such as discussed in~\cite{Kanemura:2013mc}), the septet obtains its vev from the mass term directly, $M^2_7 < 1$, and the masses of all of the new scalars are proportional to $v_\chi$ and $v_H$. Since the quartic couplings are bounded by unitarity and the vevs are fixed by $m_W$, this forces the septet masses to be significantly lighter than our DY search window, $m_7 < 4$ TeV, and so would be ruled out. This behaviour is very similar to the generalised Georgi-Machacek models without decoupling limits.

\begin{figure}[t]
    \centering
    \includegraphics[width=\textwidth]{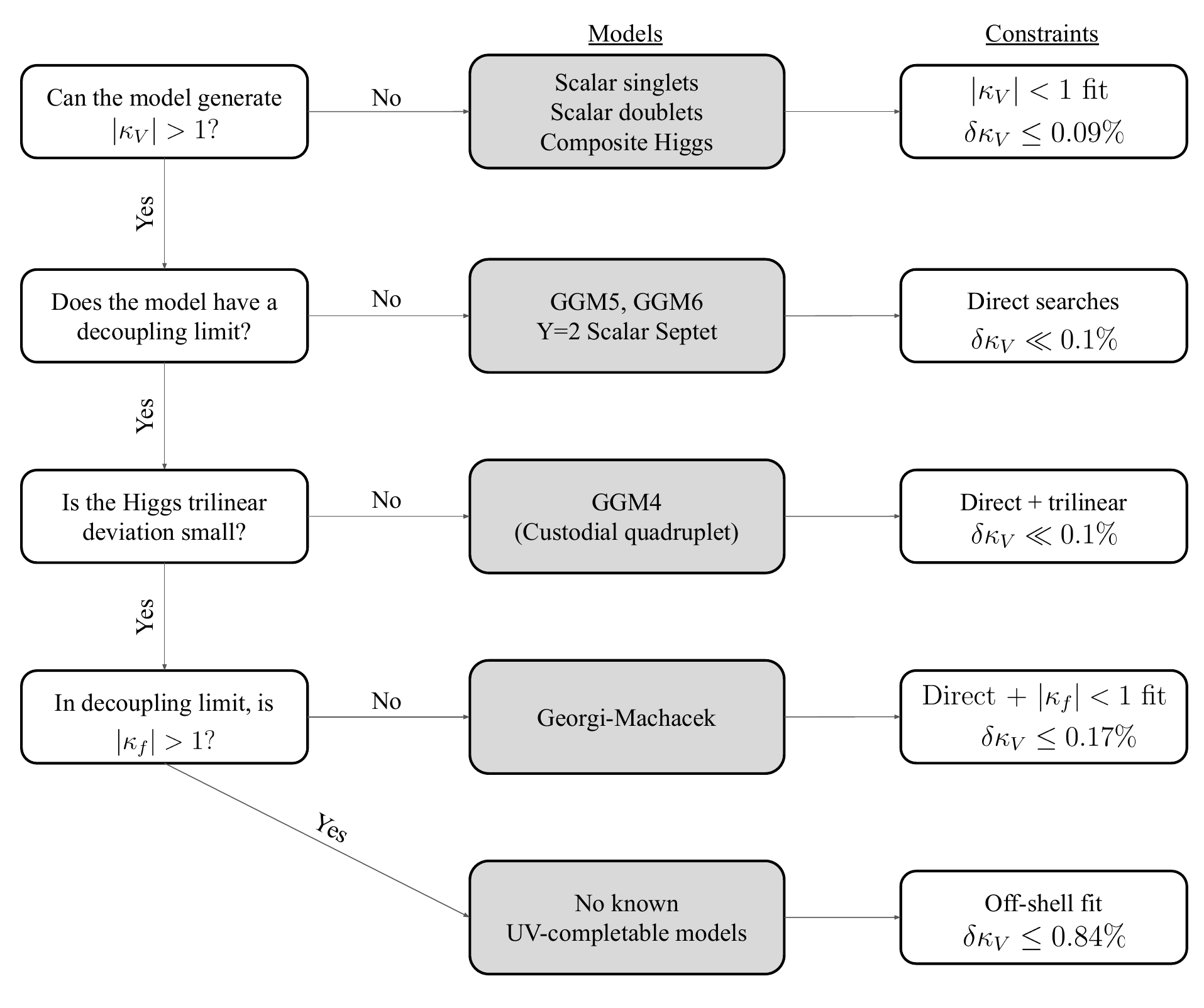}
    \caption{A summary of the projected $\kappa_V$ precision for custodially symmetric models modifying EWSB at a 10 TeV muon collider in isolation. We only show the constraints considered in this paper in the third column; additional constraints may strengthen the bounds further. The GGM models are the generalized Georgi-Machacek models presented in~\cite{Logan:2015xpa}.}
    \label{fig:ModelSummary}
\end{figure}

Before moving on, we should point out that we have not made use of the loop couplings $\kappa_\gamma$, $\kappa_{Z\gamma},$ and $\kappa_g$ in any of this discussion. For $\kappa_\gamma$ and $\kappa_{Z\gamma}$ to not exhibit observable deviations, there generically may need to be some fine tuning of the scalar quartic couplings to get the proper contribution from the charged scalars running in the loops. This is especially true for maintaining a flat direction, $\kappa_\gamma = \kappa_V$. Models surviving the combination of all of these constraints will be quite rare. To summarize, we show a flow chart in Figure~\ref{fig:ModelSummary} for models that modify EWSB and satisfy electroweak precision constraints. To consider Higgs coupling precision at a future muon collider, one has to include both the low energy Higgs coupling measurements as well as direct searches to form an accurate picture.

\section{Directly constraining \texorpdfstring{$BR_{inv}$}{BRinv}}\label{sec:BRinv}

The second requirement for a flat direction, $BR_{BSM}>0$, can likewise be constrained. As we approach decoupling in models with $|\kappa_V| > 1$, there are no light states that could be candidates for a $BR_{BSM}$, and so the theory must be supplemented by something else. Since whatever we add cannot generate the necessary $\kappa_V$, it must be fine tuned to produce a flat direction. For example, consider one of the simplest benchmark models~\cite{Cepeda:2021rql}, where the Higgs couples to a scalar singlet with a $\mathbb{Z}_2$ symmetry. In this case, if we work with a model where $|\kappa_V| > 1$ and add such a singlet to live in the flat direction by generating a $BR_{BSM}$, one would have to finely tune the cross-quartic between the Higgs and the singlet that determines $BR_{BSM}$ to match the new total Higgs width with the model's $\kappa_V$.  Such a model could manifest itself as either invisible decays or, depending on generalizations of it, as a more exotic Higgs decay, any of which may be searched for. One can of course have other scenarios where the Higgs interacts with axion-like particles, dark $U(1)$ gauge bosons, new fermions, etc to generate a $BR_{BSM}$. However, since we are adding such particles to generate a $BR_{BSM}$ independently of the model generating $|\kappa_V| > 1$, they must always be finely tuned. This tuning of independent sectors could even be more exacerbated if you consider that depending on the portal it could in principle {\em reduce} $|\kappa_V|$, making the balancing act even more difficult.  Therefore in this section we do not even consider whether there {\em exists} a complete model living in the flat direction and how robust its parameter space is, just the ability of the muon collider to test the new decay modes.

Let us consider the simplest case of a fully invisible BSM Higgs decay in more detail. This can be constrained by searching for excesses in Higgs production channels where there are associated particles to tag on. In the dominant VBF production mode, this is only possible for the $ZZ$ fusion process, since the $W^+W^-$ fusion process only has associated neutrinos. However, the forward muons in $ZZ$ fusion are highly boosted, peaking at $|\eta|\approx 5$ at a 10 TeV collider, making forward muon tagging capabilities up to high $\eta$ a requirement to use the channel. The capabilities and limitations of such a detector are not yet fully understood, although the potential of this channel for constraining $BR_{inv}$ for a variety of detector parameters was recently studied in~\cite{Ruhdorfer:2023uea}.

We first perform a sensitivity estimate of the $ZZ$ fusion process for constraining $BR_{inv}$ by looking for events that have two forward muons and missing energy, with no other particles in the event. We assume a $95\%$ efficiency for our $p_T$ range and consider a variety of energy resolutions and maximum $\eta$ reaches.  Realistically, using current Micromegas spatial resolution and a forward detector with a few T magnetic field, a resolution of ~$25\%$ seems possible for 5 TeV muons.  In principle one could use a silicon based tracker or higher magnetic field to improve the resolution, but this requires a full simulation to understand in detail so we show multiple resolutions to guide detector design targets\footnote{We thank Federico Meloni for discussions on this point and providing preliminary resolution estimates.}. For energy resolutions better than $\sim10\%$, backgrounds are dominated by the processes $\mu^+\mu^- \rightarrow \mu^+\mu^-\nu_\ell \bar{\nu}_\ell$ and $\mu^+ \mu^- \rightarrow \mu^+\mu^- \gamma$, where the associated $\gamma$ has $\eta_\gamma > 2.5$, escaping undetected down the beampipe. We apply the analysis cuts
\begin{multicols}{2}
\begin{itemize}
    \item $p_{T,{\mu^\pm}} > 20$ GeV,
    \item $|\vec{p}_{T,{\mu^+}} + \vec{p}_{T,{\mu^-}}| > 100$ GeV,
    \item $\Delta R_{\mu^+ \mu^-} > 7(9)$,
    \item $m_{\mu^+ \mu^-} > 2700(9000)$ GeV,
    \item $2.0(2.5) < |\eta_{\mu^\pm}| < \eta_{max}$,
    \item[]
\end{itemize}
\end{multicols}
\noindent where the values in parenthesis were changed going from 3 to 10 TeV. These cuts remove the overwhelming majority of those mentioned above, as well as removing any residual $\mu^+\mu^-\rightarrow \mu^+\mu^-$ events for $\sim 10\%$ energy resolution. For energy resolutions worse than this, some amount of $\mu^+\mu^-\rightarrow \mu^+\mu^-$ events begin to leak in. To loosely optimise the sensitivity, we tighten the above $|\vec{p}_{T,{\mu^+}} + \vec{p}_{T,{\mu^-}}|$ cut to $110$ GeV for 15\% resolution, tightened further to 140 GeV for 20\% and 25\% resolution and to 160 GeV for worse resolutions, although the resulting precision is still unavoidably significantly worsened by this extra background component. We have also considered modifications to the other cuts, but find that changing them does not impact the sensitivity nearly as much as the $|\vec{p}_{T,{\mu^+}} + \vec{p}_{T,{\mu^-}}|$ cut and so we leave them fixed for simplicity.

\begin{figure}[t]
    \centering
    \includegraphics[width=\textwidth]{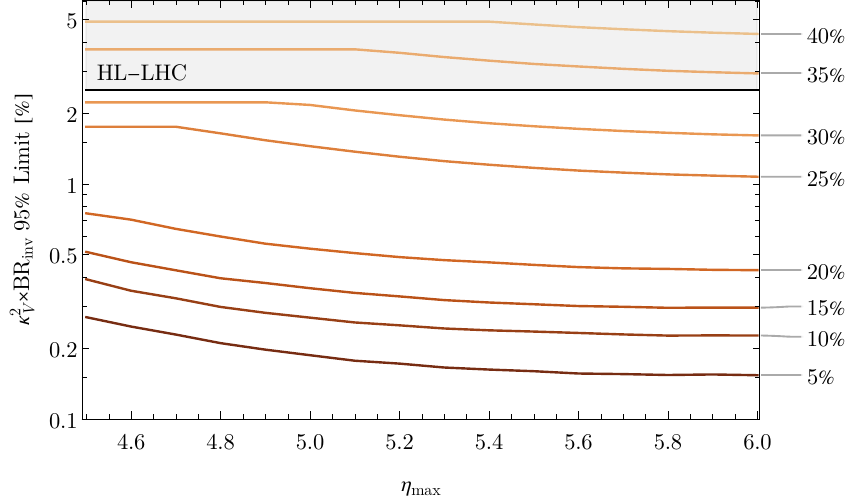}
    \caption{The estimated 95\% constraint on $\kappa_V^2 \times BR_{inv}$ from the $ZZ$ fusion $H\rightarrow inv$ search at 10 TeV as a function of the maximum forward muon detector reach, $\eta_{max}$, for a variety of energy resolution benchmarks. We have assumed $\kappa_Z = \kappa_W$ in order to show the HL-LHC projected sensitivity~\cite{Cepeda:2019klc} of $2.5\%$ for comparison.}
    \label{fig:zzf-brinv}
\end{figure}

The resulting 95\% confidence limit on $\kappa_Z^2 \times BR_{inv}$ as a function of $\eta_{max}$ is shown in Figure~\ref{fig:zzf-brinv} for various energy resolutions for a 10 TeV collider. The values at 5 and 10\% resolution we find are about a factor of two worse than the results found in~\cite{Ruhdorfer:2023uea}, while the rest of our working points are at worse resolutions than they show to demonstrate the impact of the $\mu^+\mu^-\rightarrow \mu^+\mu^-$ leaking in. In particular, it is clear that as the resolution gets worse than $\sim 15\%$, the $BR_{inv}$ reach rapidly deteriorates, which is unlikely to be improved much with a more sophisticated analysis. Detector design efforts should therefore aim at attaining a resolution better than $\sim$10-20\% for forward muon momenta of $\sim 5$ TeV if the detector is to be useful for this kind of analysis, especially since BIB effects would only introduce even more background. For the 10\% benchmark, we find a 95\% upper limit on $\kappa_Z^2 \times BR_{inv}$ of 0.22\% at 10 TeV and 1.1\% at 3 TeV. The much lower forward muon energies and pseudorapidities at 3 TeV make a 10$\%$ energy resolution much more feasible, so we do not consider $\delta E$ variations at that working point.

\begin{table}[t]
    \centering
    \renewcommand{\arraystretch}{1.3}
    \begin{tabular}{|c|c|c|c|}\hline
        Process & $\sigma$ (fb) & $A\times\epsilon$ & Number of events  \\ \hline \hline
       $(\nu_\mu \bar{\nu}_\mu)  \gamma H, \ H\rightarrow inv$ & $33\times (\kappa_W^2\textrm{BR}_{inv})$ & 0.53 & $(175 \times 10^{3})(\kappa_W^2\textrm{BR}_{inv})$\\
       $(\mu^+ \mu^-,\nu_\mu \bar{\nu}_\mu) \gamma $ & 2073 & 0.69 & $14 \times 10^6$ \\
       $(\nu_\mu \bar{\nu}_\mu) \gamma Z, \ Z\rightarrow \nu_\ell \bar{\nu}_\ell$ & 38 & 0.74 & $283 \times 10^3$ \\
      $(\nu_\mu \mu^\mp) \gamma W^\pm, \ W^\pm\rightarrow \ell^\pm \nu_\ell$& 85 & 0.17 & $146 \times 10^3$ \\ \hline
       $(\nu_\mu \bar{\nu}_\mu)  Z H, \ ZH\rightarrow \ell^+\ell^- + inv$ & $2.4 \times f_{ZH}$$BR_{inv}$ & 0.35 & $(8.5 \times 10^{3}) f_{ZH}$$BR_{inv}$\\
       $\nu_\ell\bar{\nu}_\ell\ell^+\ell^-$ & 1577 & 0.062 & $971 \times 10^3$\\
        $(\nu_\mu \bar{\nu}_\mu) ZZ, \ ZZ \rightarrow \ell^+\ell^- \nu_\ell \bar{\nu}_\ell$ & 5.7 & 0.40 & $23\times 10^3$ \\
        $(\nu_\mu \mu^\mp) W^\pm Z, \ W^\pm Z \rightarrow \ell^\pm \nu_\ell \ell^+\ell^-$ & 6.7 & 0.13 & $8.5 \times 10^3$ \\ \hline 
        $(\nu_\mu \mu^\mp) W^\pm H, \ W^\pm H \rightarrow \ell^\pm \nu_\ell + inv$ & $16\times f_{WH}$$BR_{inv}$ & 0.36 & $(56\times 10^3)f_{WH}$$BR_{inv}$ \\
        $\nu_\ell \bar{\nu}_\ell \ell^+\ell^-$ & 2820 & 0.39 & $11\times 10^6$\\\hline
        $(\nu_\mu\mu^\pm,\nu_\mu \bar{\nu}_\mu)VH, \ VH \rightarrow jj + inv$ & $72\times f_{Had}$$BR_{inv}$ & 0.16 & $(112\times 10^3)f_{Had}$$BR_{inv}$ \\
        $(\nu_\ell \bar{\nu}_\ell, \ell^\pm \nu_\ell, \ell^+ \ell^-)jj$ & 7556 & 0.17 & $13.2\times 10^6$\\
        $(\nu_\mu\bar{\nu}_\mu,\mu^+\mu^-)H, \  H\rightarrow (WW^*,ZZ^*)$ & 223 & 0.086 & $191\times 10^3$ \\
        VBF $VV \rightarrow jjjj$ & 802 & 0.18 & $1.47\times 10^6$\\
        VBF $W^\pm V \rightarrow \ell^\pm \nu_\ell jj$ & 400 & 0.067 & $267\times 10^3$\\
        VBF $ZV \rightarrow \nu_\ell \bar{\nu}_\ell jj$ & 116 & 0.26 & $306\times 10^3$\\\hline
    \end{tabular}
    \caption{Cross sections, efficiencies, and numbers of events for all signals and backgrounds for $BR_{inv}$ searches for all $\gamma H/Z H/ W^\pm H$ channels after cuts listed in the text. Here VBF refers to the combined contributions of all relevant $ZZ$ fusion, $W^+W^-$ fusion, and $W^\pm Z$ fusion processes, where the full $2\rightarrow 6$ processes with associated $(\bar{\nu}_\mu\nu_\mu,\nu_\mu\mu^\pm,\mu^+\mu^-)$ were generated.}
    \label{tab:VH}
\end{table}

Given the above caveats, we wish to do this same type of search in channels where we can tag particles in the central region, especially at 10 TeV. For all channels, the 3 TeV numbers are not very competitive, and so we will neglect their discussion, though we include the results in Table~\ref{tab:BRsummary} and the fit results including them in Appendix~\ref{app:fittabs}. We will start with $(\bar{\nu}_\mu \nu_\mu)H\gamma$, where we only tag on the photon. The $ZZ$ fusion process is found to be completely irrelevant numerically and is not considered. Since there is only one particle in the final state, there is little optimisation to be done. We choose the cuts $p_{T,\gamma}>40$ GeV and $|\eta_\ell|<2.5$, where the $p_T$ cut is chosen to be conservatively high, as BIB generates many low-$p_T$ photons. The $2\sigma$ constraint on $\kappa^2_W \times BR_{inv}$ from this channel is 4.4\%. 

The other processes to look at are the associated production modes $(\mu^\mp\nu)W^\pm H$ and $(\mu^+\mu^-,\bar{\nu}_\mu\nu_\mu)ZH$. Without assuming any forward tagging, we have three final states to look at: dilepton from $Z\rightarrow \ell^+\ell^-$, monolepton arising from $W^\pm \rightarrow \ell^\pm \nu_\ell$, and the combined hadronic channel with $Z,W^\pm\rightarrow jj$. We will look at them in order. For the dilepton final state, we can reconstruct the $Z$, allowing us to further eliminate photon backgrounds which cluster near low dilepton invariant masses. We therefore choose the looser cut $p_{T,\ell}>20$ GeV, along with the same $|\eta_{\ell}|<2.5$. We further impose $80<m_{\ell\ell}<100$ GeV and $\Delta R_{\ell\ell} > 0.2$. This channel alone yields a $2\sigma$ constraint of 23\%. For the monolepton channel, while the signal has an order of magnitude larger cross section compared to the dilepton channel, the backgrounds are also much larger than in the previous case. We only consider the background $\nu_\ell \bar{\nu}_\ell \ell^+ \ell^-$, where the dominant contributions are from $W^+W^-$ fusion $Z\rightarrow \ell^+\ell^-$ and $W^\pm Z$ fusion $W^\pm\rightarrow \ell^\pm \nu_\ell$. The total $2\sigma$ constraint is 12\% from this channel at 10 TeV. It is important to note that these constraints are not just on $BR_{inv}$, but rather a combination $f_i(\kappa_W,\kappa_Z)$$BR_{inv}$, where $f_i$ is a process dependent function of the form $f_i= a\kappa_W^2 + b\kappa_Z^2 + c\kappa_W \kappa_Z$ that includes (large) interference pieces. To determine what this function is, we scan over various $(\kappa_W,\kappa_Z)$ values and perform a fit for each analysis channel individually.

\begin{table}[t]
    \centering
    \renewcommand{\arraystretch}{1.3}
    \begin{tabular}{|c|c|c|c|}\hline
        & Process & Constrained combination $(f_i)$ & $2\sigma$ cstr. \\ \hline \hline
        \multirow{6}{*}{3 TeV} & $\mu^+\mu^- H$, $\delta E = 10\%$ & $\kappa_Z^2$$BR_{inv}$ & 1.1\%\\
        & $(\nu_\mu \bar{\nu}_\mu) H \gamma$ & $\kappa_W^2$$BR_{inv}$ & 29\%\\
        & $(\nu_\ell \bar{\nu}_\ell)ZH,\ Z\rightarrow \ell^+\ell^-$ & $(1.7 \kappa_W^2 + 1.1 \kappa_Z^2 - 1.8\kappa_W \kappa_Z)$$BR_{inv}$ & 280\%\\
        & $(\nu_\mu\mu^\mp)W^\pm H,\ W^\pm \rightarrow \ell^\pm \nu_\ell$ &  $(2.5\kappa_W^2 + 1.9 \kappa_Z^2 - 3.4 \kappa_W \kappa_Z)$$BR_{inv}$ & 91\% \\
        & $(\nu_\mu\mu^\pm,\nu_\mu \bar{\nu}_\mu)VH, \ V \rightarrow jj$ & $(2.2 \kappa_W^2  + 1.6 \kappa_Z^2 - 2.8 \kappa_W \kappa_Z)$$BR_{inv}$ & 61\% \\\hline 
        \multirow{6}{*}{10 TeV} & $\mu^+\mu^- H$, $\delta E = 10\%$ & $\kappa_Z^2$$BR_{inv}$ & 0.22\%\\
         & $(\nu_\mu \bar{\nu}_\mu) H \gamma$ & $\kappa_W^2$$BR_{inv}$ & 4.4\%\\
         & $(\nu_\ell \bar{\nu}_\ell)ZH,\ Z\rightarrow \ell^+\ell^-$ & $(4.5 \kappa_W^2 + 3.8\kappa_Z^2 - 7.3 \kappa_W \kappa_Z)$$BR_{inv}$ & 23\%\\
         &$(\nu_\mu\mu^\mp)W^\pm H,\ W^\pm \rightarrow \ell^\pm \nu_\ell$ & $(8.5 \kappa_W^2 +7.8\kappa_Z^2 - 15.3 \kappa_W \kappa_Z)$$BR_{inv}$ & 12\% \\
         &$(\nu_\mu\mu^\pm,\nu_\mu \bar{\nu}_\mu)VH, \ V \rightarrow jj$ & $(6.8 \kappa_W^2 + 6.1 \kappa_Z^2 - 11.9 \kappa_W \kappa_Z)$$BR_{inv}$ & 7.0\% \\  \hline 
         \end{tabular}
    \caption{A summary of the constraints presented in section~\ref{sec:BRinv} from $BR_{inv}$ searches. The $ZZ$ fusion numbers use a forward muon energy resolution of 10\%.}
    \label{tab:BRsummary}
\end{table} 

The hadronic channels have the additional complication of jet reconstruction, which lowers the energy resolution and smears the $W^\pm$ and $Z$ peaks. This difficulty leads them to overlap significantly, so we combine the $ZH$ and $W^\pm H$ channels, as we are not using any forward tagging information and they are therefore practically indistinguishable. We use the same jet clustering as described in section~\ref{sec:offshellanalysis}, with $R=0.5$. We require two jets with $p_{T,j} > 40$ GeV and $|\eta_j|<2.5$, with a reconstructed invariant mass between $60<M_{jj}<100$. We find a $2\sigma$ limit on $f_{Had}(\kappa_W,\kappa_Z)\times$$BR_{inv}$ from this channel of 7.0\%. We note that this channel is the most prone to new uncertainties arising from showering, jet reconstruction, and BIB since it relies on hadronic decay modes. We include details for all $\gamma H/ZH/W^\pm H$ channels including signal and background cross sections, efficiencies, and numbers of events after cuts in Table~\ref{tab:VH}.

\begin{figure}[t]
    \centering
    \includegraphics[width=.495\textwidth]{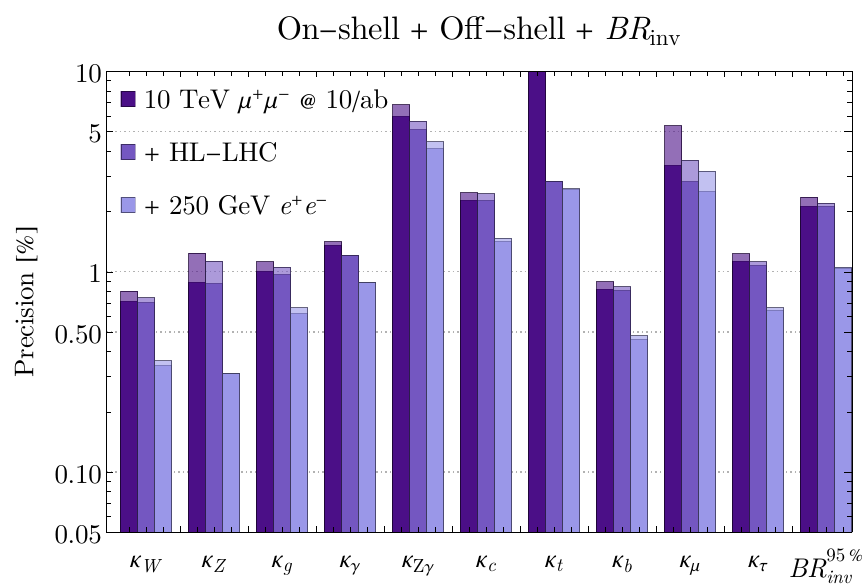}
    \includegraphics[width=.495\textwidth]{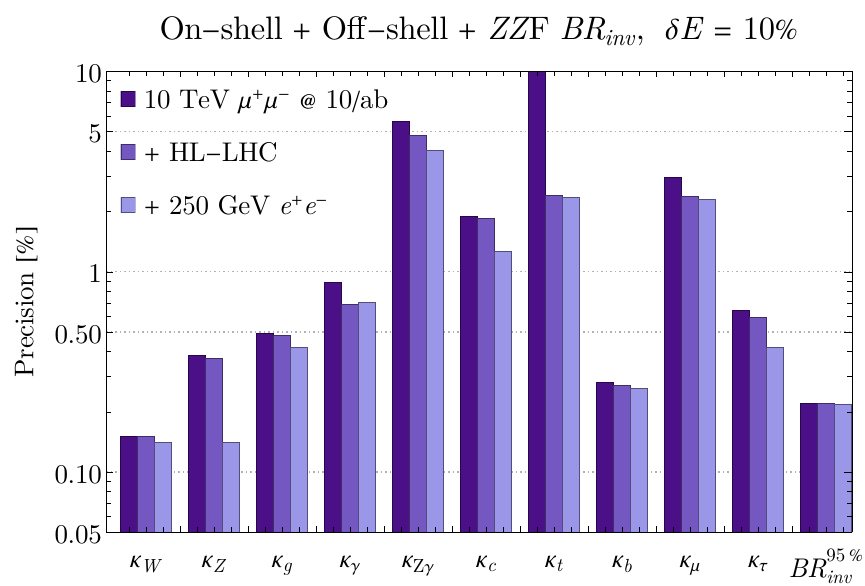}
    \caption{Fit results combining the on-shell, off-shell and $BR_{inv}$ searches assuming no exotic $BR_{BSM}$ beyond an invisible component for the 10 TeV muon collider alone and in combination with the HL-LHC and a 250 GeV $e^+e^-$ collider, where we use the fitting procedure described in Appendix~\ref{app:fits}. The left plot shows results without using the forward detector to constrain $BR_{inv}$ through $ZZ$ fusion, while the right shows the improvement using this extra channel with a forward detector muon energy resolution of 10\%. The transparent bars show the effect of removing forward tagging (see Appendix~\ref{app:methods}).}
    \label{fig:BRinvFits}
\end{figure}

A summary of the direct $BR_{inv}$ constraints is shown in Table~\ref{tab:BRsummary}. The constraints at 10 TeV are significantly stronger for every process due to a combination of the larger luminosity and much larger signal cross sections. The cancellation from interference in the VBF modes is more delicate at 10 TeV as well, which further increases the sensitivity in the full fits. With these extra $BR_{inv}$ constraints, we can look at various additional fit scenarios. In Figure~\ref{fig:BRinvFits}, we show how these constraints can improve the fit at 10 TeV if one assumes that the only $BR_{BSM}$ is from invisible decays, where we still include the off-shell information.

\section{Conclusions}\label{sec:conclusion}

In this paper we have significantly expanded the understanding of how precisely the properties of the Higgs can be measured at a high energy muon collider. Previous studies had focused on how well single Higgs precision could be achieved at a 10 TeV muon collider {\em assuming} that there was no BSM decay modes contributing to the Higgs boson width~\cite{AlAli:2021let,Forslund:2022xjq}. These studies found that a precision of up to $\mathcal{O}(.1\%)$ could be achieved under this assumption. When the width assumption is relaxed, a potential flat direction emerges in fitting Higgs properties which requires {\em both} an increase in all Higgs couplings and new BSM decay mode(s) of the Higgs. An $e^+e^-$ Higgs factory with a precise inclusive coupling measurement, or a 125 GeV muon collider with a direct width measurement can close this flat direction and preserve the sensitivity previously found in~\cite{Forslund:2022xjq}.  However, as we have shown, a 10 TeV muon collider can do this independently as well.

We have demonstrated several different approaches to closing this flat direction with a 10 TeV muon collider. The first method, most similar to the method employed by the LHC, is to use off-shell Higgs production. This is a powerful method at a high energy muon collider, as there is copious $VV\rightarrow VV$ production at all $\sqrt{\hat{s}}$. The only assumption required to translate this to Higgs precision is that $g_{hVV}(m_H)\sim g_{hVV}(\sqrt{s})$. This assumption could have a loophole if there is new physics that modifies the coupling between these scales, and therefore it is treated conservatively at the LHC. However, in the low background environment of a high energy muon collider this is a self consistent assumption for measuring the $g_{hVV}$ coupling. Nevertheless, for pure Higgs precision alone it reduces the overall precision to the $\mathcal{O}(1\%)$ level.

Another direction explored was how well new BSM contributions to the Higgs width can be constrained with a 10 TeV muon collider.  A full exotic Higgs program is still an open research question; however, as a proxy we investigated Higgs to invisible decays. The precision achievable is highly dependent on how well an energy measurement of forward muons can be done. We have shown the results for a variety of energy resolution benchmarks as a function of maximum $\eta$ reach, which we hope will be of use in detector design efforts. We have likewise included the on-shell results both with and without forward tagging up to $\eta=6$ in all fits to show the effects of the forward detector from on-shell measurements. Our upper limit with an energy resolution of 10\% is $BR^{95\%}_{inv} < 0.22\%$ at a 10 TeV muon collider, which is roughly the precision necessary to completely remove the approximate flat direction (see Appendix~\ref{Appendix:correlation}) for any $BR_{exo}$ and can therefore serve as a benchmark.

What is ultimately the most powerful tool for Higgs precision at a high energy muon collider is utilizing the energy reach directly. As mentioned, the only way to reduce the interpreted Higgs precision would be to increase all single Higgs couplings while simultaneously adding new BSM decay modes in a correlated manner. This is highly non-trivial, given that UV complete extensions of the scalar sector of the SM that could modify the Higgs couplings sufficiently have signs correlated with their representations under $SU(2)$. For instance, adding new singlets or doublets to the SM would imply a modification of the Higgs gauge boson couplings, $\vert\kappa_V\vert\leq 1$. In such a model, even if there were new BSM decay modes, the precision is back to the $\mathcal{O}(.1\%)$ level with 10 TeV muon collider alone. Given that the flat direction is populated only by $\vert\kappa_V\vert\geq 1$, as discussed in Section~\ref{sec:kv}, the only models that can achieve this in a UV consistent manner are generalizations of the Georgi-Machacek model. A muon collider can test these directly, and in particular for Higgs precision the models are {\em only} viable in the decoupling limit where $\vert\kappa_f\vert\leq 1$ and the precision is again restored to $\mathcal{O}(.1\%)$.

\begin{figure}[t]
    \centering
    \includegraphics[width=\textwidth]{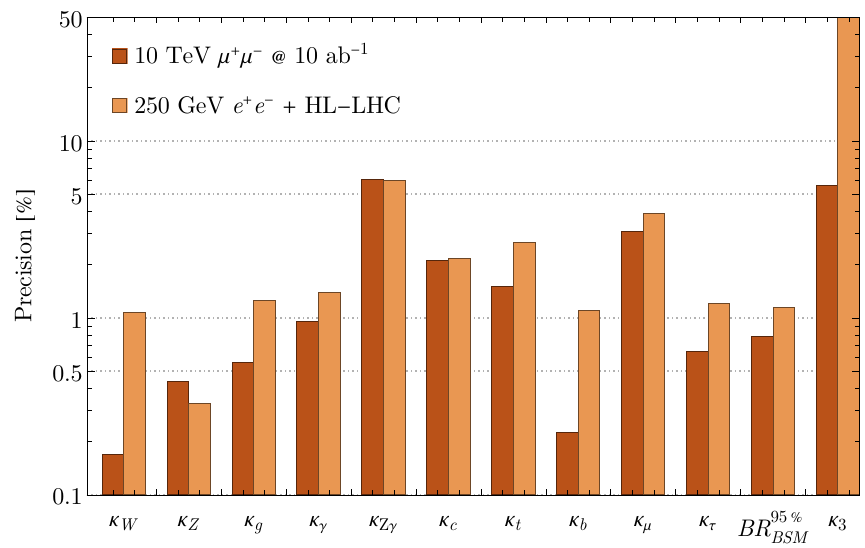}
    \caption{The current state of the art for Higgs couplings at a 10 TeV $\mu^+\mu^-$ collider in isolation compared to the combination of a 250 $e^+e^-$ collider and the HL-LHC, and we use the fitting procedure described in Appendix~\ref{app:fits}. Here $\kappa_3$ is the trilinear Higgs self-coupling result from~\cite{Buttazzo:2020uzc,Han:2020pif}. We have used the assumption $|\kappa_f| < 1$ for other couplings for the muon collider, which gives strictly weaker precisions than the assumption $|\kappa_V| < 1$ and is justified for theories violating $|\kappa_V|<1$ after incorporating direct searches at the muon collider. No assumptions are made for the $250$ GeV $e^+e^-$ + HL-LHC fit, since the direct search reach is not high enough to justify any. The muon collider fit results assume forward muon tagging up to $|\eta|\leq6$ and use the off-shell $y_t$ constraint of 1.5\% from~\cite{Liu:2023yrb,Chen:2022yiu}.}
    \label{fig:FinalFit}
\end{figure}

We have therefore demonstrated that a high energy muon collider can robustly test Higgs precision to $\mathcal{O}(.1\%)$ without having to invoke assumptions about the width. It is important to remember, of course, that single Higgs precision is not the only added benefit for Higgs physics that a muon collider allows. For example, the trilinear Higgs coupling can be measured, and there are additional observables that can test Higgs precision. As an example of this, we have included the precision achievable for a generic modification of single Higgs couplings demonstrated in this paper, as well as measurement of the triple Higgs coupling~\cite{Buttazzo:2020uzc, Han:2020pif} and a measurement of the top Yukawa using interference methods~\cite{Chen:2022yiu,Liu:2023yrb} in Figure~\ref{fig:FinalFit}.

Clearly, as shown in Figure~\ref{fig:FinalFit}, a high energy muon collider provides a striking advance for single Higgs precision, exotic branching fractions {\em and} multi Higgs tests, even if it were to be the only collider built post LHC.  If a Higgs factory is built beforehand it would add complementary knowledge.  However, by fixating on Higgs precision alone it projects our knowledge of EWSB into a lower dimensional space and does not accurately reflect the abilities of a muon collider. Obviously the true hope of any new collider is to {\em find} a deviation in the Higgs sector which could shed light on the numerous fundamental questions the Higgs has left us with. However, this means we need to understand the testable space not just in Higgs couplings, but in a UV ``model'' space as well. From this perspective we can unfold any EFT or coupling modifier prescription into a mass and coupling plane for new Higgs physics~\cite{Dawson:2022zbb,Narain:2022qud}.  A given single Higgs precision measurement lives solely on a curve in this schematic space where there could be many couplings or states.  Therefore, there are still measurements other than Higgs precision that could better test our understanding of EWSB at a muon collider, or that would be missed depending on the precision achievable in the Higgs sector.  While a complete delineation of the boundary between precision and other observables is outside the scope of this work, we can demonstrate this in the space of models that naively would cause a flat direction in Higgs precision fits, i.e. those with $\vert\kappa_V\vert\geq 1$ (generalized Georgi-Machacek models). Having a decoupling limit that could potentially avoid direct searches and severe unitarity bounds implies a tree-level coupling linear in the new heavy state, e.g. a trilinear coupling for the triplet GM model.  Therefore, despite the model having multiple parameters, we can focus on the effect of this coupling to the SM Higgs compared to the mass of the new state to illustrate the parameter space covered in different approaches.

\begin{figure}[t]
    \centering
    \includegraphics[width=\textwidth]{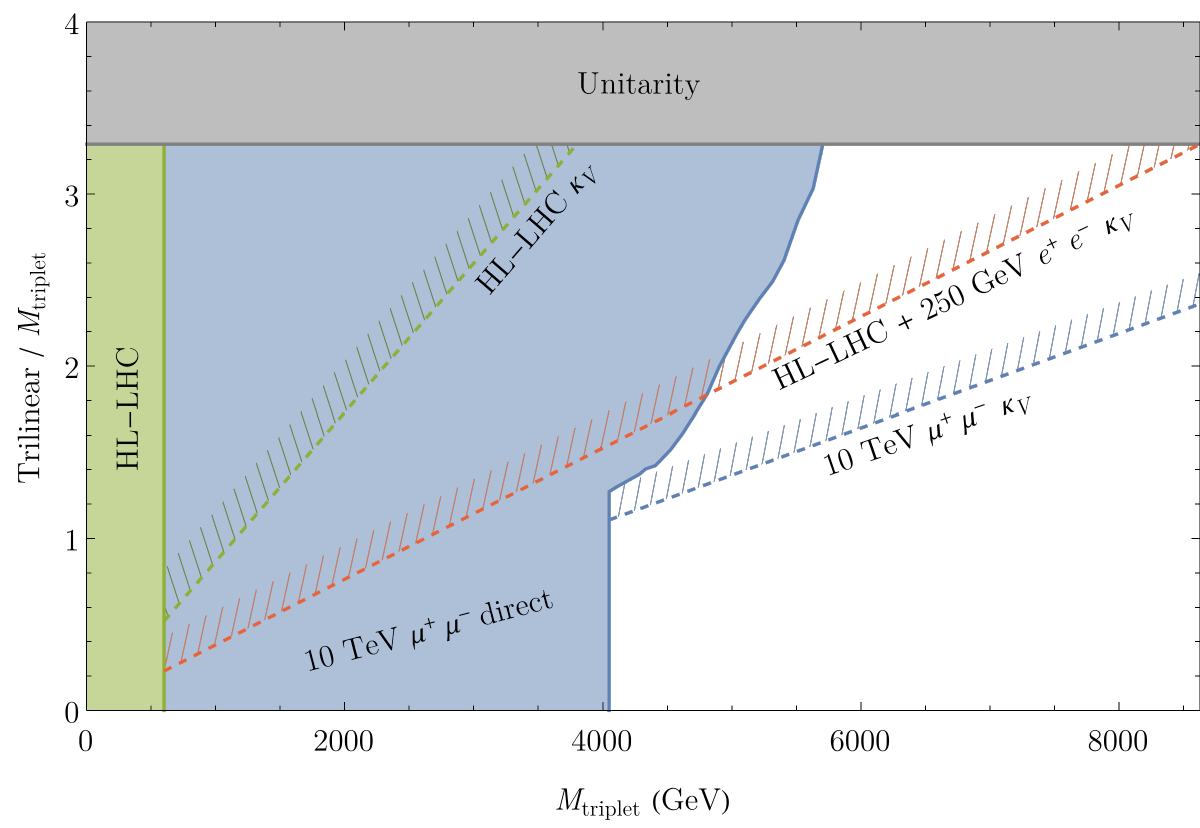}
    \caption{Illustrating the larger model space constraints on the GM model from the coupling to the SM and mass of the new states.  Here we have projected in the $(M_1/\mu_3$, $\mu_3)$ plane, where $M_1$ is the coefficient of the trilinear coupling between the SM Higgs doublet and the custodial triplet, and $\mu_3$ is approximately the triplet mass. The gray region is excluded by perturbative unitarity of $\lambda_1$, the quartic self interaction of the scalar doublet. The green and dark blue show the HL-LHC and 10 TeV $\mu^+\mu^-$ collider direct search reaches, respectively, as discussed in Section~\ref{sec:GM}. The dashed lines show the constraints from $\kappa$ precision, where the green is from the HL-LHC $BR_{BSM}=0$ fit, red is for the combination of the HL-LHC and a 250 GeV $e^+e^-$ collider in a general fit, and the blue is our muon collider fit with the $|\kappa_f|<1$ assumption, which is justified due to the direct searches removing the $\mu_3 \lesssim 4$ TeV region.}
    \label{fig:GMtril}
\end{figure}

In Figure~\ref{fig:GMtril} we show the reach of a high energy muon collider in this generic coupling versus mass plane for the GM model, where the solid blue (green) region is the union of the muon collider (HL-LHC) direct searches presented in Section~\ref{sec:GM}, the gray region is the bound from perturbative unitarity, and the dashed lines are the reach from the appropriate $\kappa$-fits for different collider options using the relations in the decoupling limit~\eqref{eq:kvDec}. The Higgs precision alone is very impressive, and a muon collider can extend beyond the LHC and future $e^+e^-$ colliders. However, what is more impressive is the ability of the muon collider to search for new physics in multiple ways in the same region of parameter space. For instance, if there is a deviation in a Higgs coupling, up to multi-TeV scale masses the muon collider can test this directly and discover the new states responsible in the same experiment. Furthermore, by realizing that Higgs physics is more than SM Higgs couplings, at smaller coupling to BSM states and ``low'' masses we see that a muon collider can discover extensions of EWSB in regions untestable through standard Higgs precision projections alone. Therefore, if a muon collider is built, it is crucial to change our paradigm of separating precision physics from other observables if one wants a complete picture of its capabilities. 

\acknowledgments

We would like to thank Dimitrios Athanasakos, Luca Giambastiani, Simone Pagan Griso, Samuel Homiller, Sergo Jindariani, Zhen Liu, Donatella Lucchesi, Federico Meloni, Lorenzo Sestini, and Mauro Valli for useful conversations and details that allowed this study to be completed. This work was supported by the National Science Foundation grant PHY-2210533. PM would also like to acknowledge the Aspen Center for Physics, which is supported by National Science Foundation grant PHY-2210452 where this work was completed.

\bibliographystyle{utphys}
\bibliography{bib}

\appendix

\section{Event generation and detector assumptions}\label{app:methods}
All event generation and detector simulation for our muon collider results is identical to our on-shell analysis~\cite{Forslund:2022xjq}, which we will briefly review here. Events are generated using {\sc MadGraph5}~\cite{Alwall:2014}, with parton showering handled by {\sc Pythia8}~\cite{Sjostrand:2014zea}. Any VBF process with associated forward muons is generated with a cut $p_{T_\mu} > 10$ GeV to regulate the phase space divergences from virtual photons. All other generation level cuts applied are strictly weaker than analysis cuts and do not change results. We do not include any contributions of collinear low-virtuality photons or the potential extra contributions from $q/g$ components of the muons~\cite{Han:2021kes,Garosi:2023bvq}. Both of these backgrounds peak at low invariant masses and are therefore much less important in our off-shell results than the on-shell results of~\cite{Forslund:2022xjq}. Similarly, since we consider final states with two vector bosons $W^\pm$ or $Z$, our invariant mass cuts when reconstruct them will remove most of these backgrounds. We take $m_H = 125$ GeV, $\Gamma^{SM}_H = 4.07$ MeV, and use the SM branching ratios from the CERN yellow report~\cite{LHCHiggsCrossSectionWorkingGroup:2013rie}.

Detector simulation is handled using {\sc Delphes}~\cite{deFavereau:2013fsa} fast detector simulation with the muon detector card taken to be mostly a hybrid of FCC-hh and CLIC cards for efficiencies and energy resolution. The card has a strict $\eta$ cutoff of $|\eta|<2.5$ for all detected particles to approximate the effect of a tungsten nozzle leading to the interaction point~\cite{Foster:1995ru,Bartosik:2019dzq,Mokhov:2011zzb}, which has been found to be necessary for mitigating BIB to a manageable level in full simulation studies at 1.5 TeV. Given that the BIB is peaked more in the forward direction as the energy goes higher, this $\eta$ cut is quite conservative and may be loosened in future full simulation studies. This {\sc Delphes} card is clearly just an approximation for a realistic detector environment and does not constitute any final say. 

Beyond the $|\eta|<2.5$ cut, a hypothetical forward muon detector is included in the {\sc Delphes} card with an efficiency of 90(95)\% for muons with $0.5<p_T<1$ GeV and $p_T > 1$ GeV, respectively. The achievable efficiency and resolution of such a detector are a topic of active research, so we include results both with and without this forward muon detector extending up to $|\eta|<6$ for on-shell results\footnote{The on-shell results with forward tagging do not use any forward muon energy measurements, only the efficiency.}, and we do not use it at all for our off-shell results in section~\ref{sec:offshellanalysis}. For the $ZZ$ fusion $BR_{inv}$ search in section~\ref{sec:BRinv}, we consider a variety of potential energy efficiencies and maximum $\eta$ reaches of such a detector in a similar manner to~\cite{Ruhdorfer:2023uea}. 

For hadronic final states, we use the exclusive Valencia jet reconstruction algorithm, a generalization of longitudinally invariant $e^+e^-$ collider algorithms that performs well in the presence of forward peaked $\gamma\gamma\rightarrow hadrons$ backgrounds in simulations~\cite{Boronat:2014hva,Boronat:2016tgd}. We set $\gamma=\beta=1$ and $R=0.5$ for new results presented here, although some on-shell results used different $R$ values~\cite{Forslund:2022xjq}. 

\section{Fitting procedure}\label{app:fits}

For all fits presented in this paper, we implement our observables in the Bayesian analysis framework HEPfit~\cite{DeBlas:2019ehy}. We choose flat priors for all parameters with ranges large enough to not affect our posterior distributions. Input observables are taken to be the standard model expectation with Gaussian errors. While our on-shell results usually yield Gaussian posterior distributions in accordance with the simple analysis performed in~\cite{Forslund:2022xjq}, inclusion of the off-shell and $BR_{inv}$ constraints in sections~\ref{sec:offshellanalysis} and~\ref{sec:BRinv} makes a more general framework necessary. All precisions we give are the difference between the standard model prediction and the furthest 68\% confidence interval of the marginalised posterior distribution, which closely reproduce the Gaussian expectation for the $\kappa$-0 and $|\kappa_V|\leq 1$ fits while avoiding overestimating sensitivities in the general fits where there is a large asymmetry from an approximate flat direction.

We show fits for the muon collider in isolation, in combination with the HL-LHC, and in combination with both the LHC and a 250 GeV $e^+e^-$ collider, which we take to be CEPC. The HL-LHC inputs are taken from~\cite{Cepeda:2019klc} assuming S2 systematics, and the CEPC inputs are taken to be those in~\cite{An:2018dwb}, where we include the relevant correlation matrices for both. Of course, the results would be similar for any comparable $e^+e^-$ collider such as the 250 GeV stages of ILC~\cite{ILCInternationalDevelopmentTeam:2022izu}, FCC-$ee$~\cite{Bernardi:2022hny} or C$^3$~\cite{Bai:2021rdg}, but a comparison of all of these akin to~\cite{deBlas:2019rxi} is beyond the scope of this paper. Results including a 125 GeV $\mu^+\mu^-$ collider use the inputs from~\cite{deBlas:2022aow}. We neglect all uncertainties on other SM parameters such as $m_t$, assuming they will be under control and subdominant by the time a future muon collider is finished running. 

In~\cite{Chen:2022yiu,Liu:2023yrb}, the capabilities of off-shell methods at a muon collider for a measurement of the top Yukawa coupling were explored, giving a precision of 1.5\% including the semileptonic and hadronic channels. This measurement does not use the same methodology as we have- in particular, they work with the top quarks without simulating their decays and do not perform any fast detector simulation. However, since this number is far better than our on-shell $\kappa_t$ measurement from $t\bar{t}H$ production~\cite{Forslund:2022xjq}, it is still important to consider its effect on our fits. While $1.5\%$ is better than our on-shell number and even better than the HL-LHC projection, it is still subdominant to our measurements on $\kappa_V$ and $\kappa_b$. This implies that even in the general fits, the improvement from this measurement is nearly entirely on $\kappa_t$ itself, and doesn't modify our other precisions in any meaningful way. We have performed the fits including this constraint as an absolute measurement on $y_t$, and find results consistent with this intuition. In particular, we find that regardless of our assumptions, $\delta\kappa_t = 1.5\%$ for the fits with the muon collider in isolation. In combination with the HL-LHC, this improves to $\delta\kappa_t = 1.2\%$ for the fits using assumptions to break the degeneracy, while it is a slightly worse 1.4\% for the fully general off-shell fit. Additionally combining with a 250 GeV $e^+e^-$ collider improves the latter of these numbers to $1.3\%$ without changing the former.

\section{Correlations}\label{Appendix:correlation}
\begin{figure}[t]
    \centering
    \includegraphics[width=0.495\textwidth]{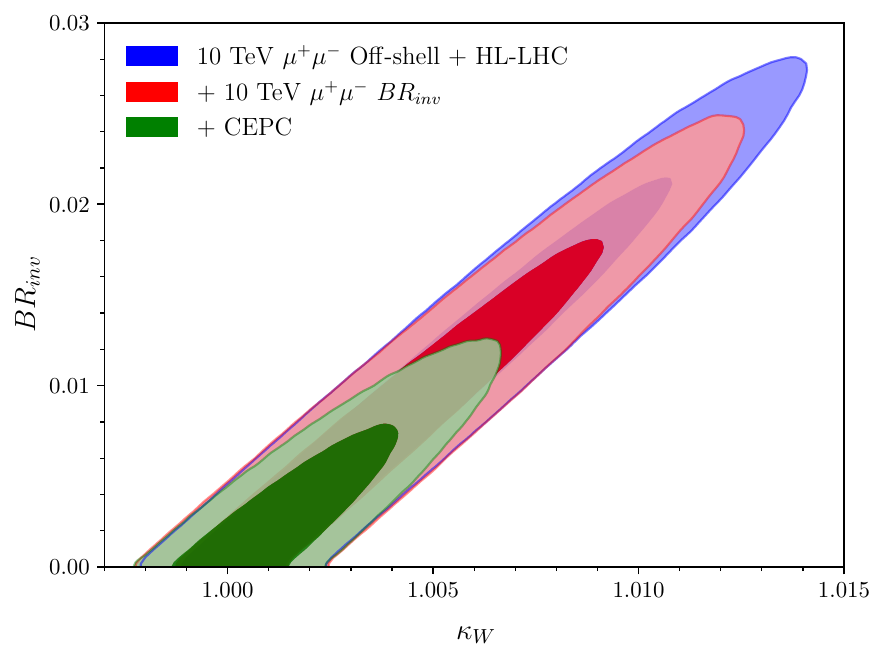}
    \includegraphics[width=0.495\textwidth]{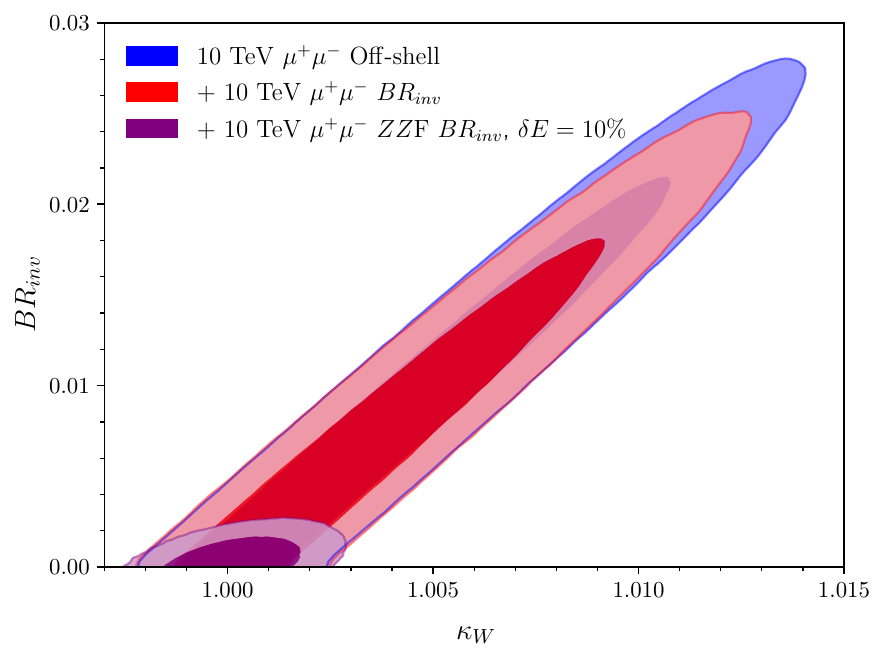}
    \caption{Correlations between $\kappa_W$ and $BR_{inv}$ for our fits including on-shell and off-shell information, assuming all $BR_{BSM} = BR_{inv}$. The filled and transparent contours show the 68\% and 95\% confidence intervals, respectively. The left plot shows results for the 10 TeV $\mu^+\mu^-$ collider in combination with the HL-LHC, the improvement when adding $BR_{inv}$ searches, and when additionally combined with a 250 GeV $e^+e^-$ collider. The right plot shows results for the 10 TeV $\mu^+\mu^-$ collider alone, the improvement from $BR_{inv}$ searches, and the addition of the $ZZ$F $BR_{inv}$ search constraint discussed in section~\ref{sec:BRinv} assuming a forward muon energy resolution of 10\%.}
    \label{fig:correlations}
\end{figure}
In contrast to $e^+e^-$ colliders, the measurements breaking our degeneracy at a muon collider have a worse precision than many of the on-shell constraints. This causes the flat direction to persist in the form of a very strong correlation between the accurately measured $\kappa$'s and $BR_{inv}$. This is most clearly visible for $\kappa_W$. The allowed region in the $(\kappa_W,BR_{inv})$ plane is shown in Figure~\ref{fig:correlations} for a variety of our 10 TeV muon collider fit scenarios. The left plot shows (in combination with the HL-LHC) the off-shell fit, the improvement when $BR_{inv}$ searches are included, and in combination with a 250 GeV $e^+e^-$ collider. The right plot shows results for the muon collider alone for only off-shell, with $BR_{inv}$ searches, and with the $ZZ$ fusion $BR_{inv}$ search with a forward muon energy resolution of $10\%$. The correlations between $BR_{inv}$ and $\kappa_b$ or $\kappa_Z$ are very similar. 

\section{Tabulated fit results}\label{app:fittabs}
For convenience, here we include tables with most of the muon collider fit results throughout the paper and the equivalents at 3 TeV. All fits use the methodology in Appendix~\ref{app:fits}. The fits with $BR_{inv}=0$ are slightly different than in~\cite{Forslund:2022xjq} due to using HEPfit and slightly different CEPC inputs, so we include them here as well. Tables~\ref{K0tab} and~\ref{KvLT1tab} show the fit results for the left and right charts in Figure~\ref{fig:K0fits}, respectively; Tables~\ref{K0tab3} and~\ref{KvLT1tab3} show the equivalent fits for 3 TeV. Tables~\ref{KfTab3} and~\ref{KfTab10} show results for on-shell fits with the assumption $|\kappa_f| < 1$ to break the degeneracy. Tables~\ref{KOffshTab3} and~\ref{KOffshTab10} show the fit results at 3 TeV and 10 TeV when using the off-shell information to remove the flat direction, where the latter is the same as the left chart of Figure~\ref{fig:KOffshFits}. Tables~\ref{KBRinv3tab} and~\ref{KBRinv10tab} show the improvement over the off-shell fits when incorporating the $H\gamma$, $HZ$, and $HW^\pm$ $BR_{inv}$ searches discussed in Section~\ref{sec:BRinv} under the assumption that the only BSM Higgs branching fraction is to invisibles. The results in Table~\ref{KBRinv10tab} correspond to those in the left plot of Figure~\ref{fig:BRinvFits}. Finally, the improvement over these fits from a forward muon detector up to $|\eta|=6$ with an optimistic energy resolution of $\delta E=10\%$ is shown in Table~\ref{KZZFtab}, where the 10 TeV numbers correspond to those shown in the right chart of Figure~\ref{fig:BRinvFits}.

\begin{table}[t]
\renewcommand{\arraystretch}{1.5}
\setlength{\arrayrulewidth}{.3mm}
\setlength{\tabcolsep}{1 em}
\begin{center}
\begin{tabular}{|c|c cc|ccc|}
\multicolumn{7}{c}{3 TeV $\mu^+\mu^-$ Collider $BR_{BSM} = 0$ Fit Result [\%]} \\ \hline
 & \multicolumn{3}{c|}{With Forward Tagging} & \multicolumn{3}{c|}{Without Forward Tagging} \\ 
 & $\mu^+\mu^-$ & +HL-LHC & +CEPC  & $\mu^+\mu^-$ & +HL-LHC & +CEPC  \\ 
\hline
$\kappa_W$ & 0.38 & 0.36 & 0.31 & 0.56 & 0.39 & 0.33 \\
$\kappa_Z$ & 1.2 & 0.80 & 0.12 & 5.2 & 1.0 & 0.12 \\
$\kappa_g$ & 1.7 & 1.1 & 0.67 & 2.0 & 1.2 & 0.70 \\
$\kappa_\gamma$ & 3.2 & 1.1 & 1.0 & 3.2 & 1.1 & 1.0 \\
$\kappa_{Z\gamma}$ & 25. & 8.5 & 5.6 & 28. & 8.6 & 5.7 \\
$\kappa_c$ & 6.0 & 6.1 & 1.7 & 6.9 & 6.9 & 1.7 \\
$\kappa_t$ & 38. & 2.4 & 2.4 & 37. & 2.4 & 2.4 \\
$\kappa_b$ & 0.87 & 0.78 & 0.44 & 0.98 & 0.79 & 0.45 \\
$\kappa_\mu$ & 15. & 4.0 & 3.6 & 23. & 4.1 & 3.7 \\
$\kappa_\tau$ & 2.1 & 1.2 & 0.59 & 2.3 & 1.2 & 0.61 \\
\hline
\end{tabular}
\end{center}
\caption{Tabulated results for a 3 TeV $\mu^+\mu^-$ collider $BR_{BSM}=0$ fit.}\label{K0tab3}
\end{table}

\begin{table}[t]
\renewcommand{\arraystretch}{1.5}
\setlength{\arrayrulewidth}{.3mm}
\setlength{\tabcolsep}{1 em}
\begin{center}
\begin{tabular}{|c|c cc|ccc|}
\multicolumn{7}{c}{10 TeV $\mu^+\mu^-$ Collider $BR_{BSM} = 0$ Fit Result [\%]} \\ \hline
 & \multicolumn{3}{c|}{With Forward Tagging} & \multicolumn{3}{c|}{Without Forward Tagging} \\ 
 & $\mu^+\mu^-$ & +HL-LHC & +CEPC  & $\mu^+\mu^-$ & +HL-LHC & +CEPC  \\ 
\hline
$\kappa_W$ & 0.11 & 0.10 & 0.10 & 0.16 & 0.13 & 0.11 \\
$\kappa_Z$ & 0.35 & 0.33 & 0.10 & 1.4 & 0.8 & 0.11 \\
$\kappa_g$ & 0.46 & 0.43 & 0.38 & 0.53 & 0.49 & 0.42 \\
$\kappa_\gamma$ & 0.84 & 0.67 & 0.66 & 0.85 & 0.68 & 0.66 \\
$\kappa_{Z\gamma}$ & 5.6 & 4.8 & 4.0 & 6.6 & 5.3 & 4.4 \\
$\kappa_c$ & 1.9 & 1.8 & 1.3 & 2.1 & 2.0 & 1.3 \\
$\kappa_t$ & 12 & 2.4 & 2.4 & 12 & 2.4 & 2.4 \\
$\kappa_b$ & 0.24 & 0.23 & 0.21 & 0.26 & 0.25 & 0.22 \\
$\kappa_\mu$ & 2.9 & 2.4 & 2.3 & 5.0 & 3.2 & 2.94 \\
$\kappa_\tau$ & 0.59 & 0.54 & 0.38 & 0.64 & 0.57 & 0.41 \\
\hline
\end{tabular}
\end{center}
\caption{Tabulated results for the 10 TeV $\mu^+\mu^-$ collider $BR_{BSM}=0$ fit in Figure~\ref{fig:K0fits}.}\label{K0tab}
\end{table}

\begin{table}[t]
\renewcommand{\arraystretch}{1.5}
\setlength{\arrayrulewidth}{.3mm}
\setlength{\tabcolsep}{1 em}
\begin{center}
\begin{tabular}{|c|c cc|ccc|}
\multicolumn{7}{c}{3 TeV $\mu^+\mu^-$ Collider $|\kappa_V|< 1$ Fit Result [\%]} \\ \hline
 & \multicolumn{3}{c|}{With Forward Tagging} & \multicolumn{3}{c|}{Without Forward Tagging} \\ 
 & $\mu^+\mu^-$ & +HL-LHC & +CEPC  & $\mu^+\mu^-$ & +HL-LHC & +CEPC  \\ 
\hline
$\kappa_W$ &  0.32 & 0.31 & 0.28 & 0.30 & 0.31 & 0.30 \\
$\kappa_Z$ &  1.2 & 0.80 & 0.10 & 1.6 & 0.95 & 0.10 \\
$\kappa_g$ &  1.6 & 1.1 & 0.64 & 1.9 & 1.2 & 0.66 \\
$\kappa_\gamma$ &  3.2 & 1.1 & 1.0 & 3.2 & 1.1 & 1.0 \\
$\kappa_{Z\gamma}$ &  26. & 8.5 & 5.6 & 29. & 8.6 & 5.6 \\
$\kappa_c$ &  6.0 & 5.8 & 1.6 & 6.7 & 6.6 & 1.7 \\
$\kappa_t$ &  37. & 2.4 & 2.4 & 38. & 2.4 & 2.4 \\
$\kappa_b$  &  0.85 & 0.76 & 0.38 & 0.90 & 0.78 & 0.39 \\
$\kappa_\mu$ &  15. & 4.0 & 3.6 & 23. & 4.1 & 3.7 \\
$\kappa_\tau$ &  2.1 & 1.1 & 0.55 & 2.2 & 1.2 & 0.55 \\
$BR_{BSM}^{95\%}$ &  1.3 & 1.1 & 0.39 & 1.3 & 1.2 & 0.40 \\
\hline
\end{tabular}
\end{center}
\caption{Tabulated results for a 3 TeV $\mu^+\mu^-$ collider $|\kappa_V|<1$ fit.}\label{KvLT1tab3}
\end{table}

\begin{table}[t]
\renewcommand{\arraystretch}{1.5}
\setlength{\arrayrulewidth}{.3mm}
\setlength{\tabcolsep}{1 em}
\begin{center}
\begin{tabular}{|c|c cc|ccc|}
\multicolumn{7}{c}{10 TeV $\mu^+\mu^-$ Collider $|\kappa_V|< 1$ Fit Result [\%]} \\ \hline
 & \multicolumn{3}{c|}{With Forward Tagging} & \multicolumn{3}{c|}{Without Forward Tagging} \\ 
 & $\mu^+\mu^-$ & +HL-LHC & +CEPC  & $\mu^+\mu^-$ & +HL-LHC & +CEPC  \\ 
\hline
$\kappa_W$ & 0.09 & 0.09 & 0.08 & 0.09 & 0.09 & 0.09 \\
$\kappa_Z$ & 0.35 & 0.33 & 0.09 & 0.78 & 0.58 & 0.09 \\
$\kappa_g$ & 0.44 & 0.41 & 0.36 & 0.5 & 0.47 & 0.4 \\
$\kappa_\gamma$ & 0.84 & 0.67 & 0.66 & 0.84 & 0.68 & 0.66 \\
$\kappa_{Z\gamma}$ & 5.6 & 4.8 & 4.0 & 6.7 & 5.4 & 4.3 \\
$\kappa_c$ & 1.7 & 1.7 & 1.2 & 1.9 & 1.9 & 1.3 \\
$\kappa_t$ & 11. & 2.4 & 2.4 & 11.6 & 2.4 & 2.4 \\
$\kappa_b$  & 0.22 & 0.22 & 0.2 & 0.24 & 0.23 & 0.21 \\
$\kappa_\mu$ & 2.9 & 2.4 & 2.3 & 5.0 & 3.2 & 3.0 \\
$\kappa_\tau$ & 0.57 & 0.53 & 0.37 & 0.61 & 0.56 & 0.39 \\
$BR_{BSM}^{95\%}$ & 0.34 & 0.34 & 0.24 & 0.35 & 0.34 & 0.25 \\
\hline
\end{tabular}
\end{center}
\caption{Tabulated results for the 10 TeV $\mu^+\mu^-$ collider $|\kappa_V|<1$ fit in Figure~\ref{fig:K0fits}.}\label{KvLT1tab}
\end{table}

\begin{table}[t]
\renewcommand{\arraystretch}{1.5}
\setlength{\arrayrulewidth}{.3mm}
\setlength{\tabcolsep}{1 em}
\begin{center}
\begin{tabular}{|c|c cc|ccc|}
\multicolumn{7}{c}{3 TeV $\mu^+\mu^-$ Collider $|\kappa_f| < 1$ Fit Result [\%]} \\ \hline
 & \multicolumn{3}{c|}{With Forward Tagging} & \multicolumn{3}{c|}{Without Forward Tagging} \\ 
 & $\mu^+\mu^-$ & +HL-LHC & +CEPC  & $\mu^+\mu^-$ & +HL-LHC & +CEPC  \\ 
\hline
$\kappa_W$ &  0.64 &	0.48 &	0.38 &	0.69 &	0.52 &	0.42 \\
$\kappa_Z$ &  1.6 &	0.93 &	0.28 &	7.5 &	1.1 &	0.28 \\
$\kappa_g$ &  2.1 &	1.4 &	0.89 &	2.6 &	1.5 &	0.9 \\
$\kappa_\gamma$ &  3.7 &	1.3 &	1.1  &	3.6 &	1.3 &	1.1 \\
$\kappa_{Z\gamma}$ &  30. &	9.3 &	6.2	 &  32. &	9.2 &	6.1 \\
$\kappa_c$ &  7.2 &	7.4 &	2.0  &	8.2 &	8.7 &	2.0 \\
$\kappa_t$ &    54. &	2.5 &	2.5  &	54. &	2.5	&   2.4 \\
$\kappa_b$ &   0.82 &  0.76 &	0.47 &	0.90 &	0.82 &	0.48 \\
$\kappa_\mu$  &  18. &	4.1 &	3.7  &	29. &	4.2	&   3.8 \\
$\kappa_\tau$ &  2.4 &	1.3 &	0.67 &	2.5 &	1.26 &	0.67 \\
$BR_{BSM}^{95\%}$ & 2.9 &	2.1 &	0.74 &	3.2 &	2.2 &	0.75 \\
\hline
\end{tabular}
\end{center}
\caption{Tabulated results for a 3 TeV $\mu^+\mu^-$ collider $|\kappa_f| < 1$ fit.}\label{KfTab3}
\end{table}

\begin{table}[t]
\renewcommand{\arraystretch}{1.5}
\setlength{\arrayrulewidth}{.3mm}
\setlength{\tabcolsep}{1 em}
\begin{center}
\begin{tabular}{|c|c cc|ccc|}
\multicolumn{7}{c}{10 TeV $\mu^+\mu^-$ Collider $|\kappa_f| < 1$ Fit Result [\%]} \\ \hline
 & \multicolumn{3}{c|}{With Forward Tagging} & \multicolumn{3}{c|}{Without Forward Tagging} \\ 
 & $\mu^+\mu^-$ & +HL-LHC & +CEPC  & $\mu^+\mu^-$ & +HL-LHC & +CEPC  \\ 
\hline
$\kappa_W$ & 0.17 & 0.16 & 0.14 & 0.20 & 0.18 & 0.14 \\
$\kappa_Z$ & 0.44 & 0.41 & 0.22 & 2.1  & 1.0  & 0.23 \\
$\kappa_g$ & 0.56 & 0.56 & 0.49 & 0.67 & 0.64 & 0.54 \\
$\kappa_\gamma$ & 0.95 & 0.73 & 0.69 & 0.96 & 0.77 & 0.71 \\
$\kappa_{Z\gamma}$ & 6.1  & 5.2 & 4.3 & 7.3 & 5.8 & 5.0 \\
$\kappa_c$ & 2.1  & 2.1 & 1.5 & 2.5 & 2.5 & 1.6 \\
$\kappa_t$ & 14.  & 2.5 & 2.5 & 13. & 2.5 & 2.5 \\
$\kappa_b$ & 0.22 & 0.22 & 0.21 & 0.25 & 0.24 & 0.23 \\
$\kappa_\mu$  & 3.1  & 2.4 & 2.4 & 5.3 & 3.3 & 3.2 \\
$\kappa_\tau$ & 0.65 & 0.59 & 0.44 & 0.70 & 0.63 & 0.46 \\
$BR_{BSM}^{95\%}$ & 0.78 & 0.74 & 0.54 & 0.89 & 0.80 & 0.54 \\
\hline
\end{tabular}
\end{center}
\caption{Tabulated results for the 10 TeV $\mu^+\mu^-$ collider $|\kappa_f| < 1$ fit in Figure~\ref{fig:GMfits}.}\label{KfTab10}
\end{table}

\begin{table}[t]
\renewcommand{\arraystretch}{1.5}
\setlength{\arrayrulewidth}{.3mm}
\setlength{\tabcolsep}{1 em}
\begin{center}
\begin{tabular}{|c|c cc|ccc|}
\multicolumn{7}{c}{3 TeV $\mu^+\mu^-$ Collider Off-shell Fit Result [\%]} \\ \hline
 & \multicolumn{3}{c|}{With Forward Tagging} & \multicolumn{3}{c|}{Without Forward Tagging} \\ 
 & $\mu^+\mu^-$ & +HL-LHC & +CEPC  & $\mu^+\mu^-$ & +HL-LHC & +CEPC  \\ 
\hline
$\kappa_W$ &  5.6 & 5.7 & 0.56 & 5.9 & 5.7 & 0.58 \\
$\kappa_Z$ &  6.1 & 5.9 & 0.32 & 7.5 & 5.9 & 0.33 \\
$\kappa_g$ &  6.5 & 6.1 & 0.96 & 7.2 & 6.1 & 0.98 \\
$\kappa_\gamma$ &  7.7 & 6.1 & 1.2 & 8.0 & 6.1 & 1.3 \\
$\kappa_{Z\gamma}$ &  26. & 12. & 5.7 & 29. & 12. & 5.8 \\
$\kappa_c$ &  10. & 10. & 1.9 & 11. & 11. & 1.9 \\
$\kappa_t$ &  39. & 7.0 & 2.7 & 40. & 7.0 & 2.7 \\
$\kappa_b$ &  6.0 & 6.0 & 0.77 & 6.4 & 6.0 & 0.78 \\
$\kappa_\mu$ &  18. & 8.3 & 3.9 & 25. & 8.3 & 3.7 \\
$\kappa_\tau$ &  6.8 & 6.2 & 0.94 & 7.4 & 6.2 & 0.95 \\
$BR_{BSM}^{95\%}$ &  14. & 14. & 1.1 & 15. & 14. & 1.1 \\
\hline
\end{tabular}
\end{center}
\caption{Tabulated results for a 3 TeV $\mu^+\mu^-$ collider off-shell fit.}\label{KOffshTab3}
\end{table}

\begin{table}[t]
\renewcommand{\arraystretch}{1.5}
\setlength{\arrayrulewidth}{.3mm}
\setlength{\tabcolsep}{1 em}
\begin{center}
\begin{tabular}{|c|c cc|ccc|}
\multicolumn{7}{c}{10 TeV $\mu^+\mu^-$ Collider Off-shell Fit Result [\%]} \\ \hline
 & \multicolumn{3}{c|}{With Forward Tagging} & \multicolumn{3}{c|}{Without Forward Tagging} \\ 
 & $\mu^+\mu^-$ & +HL-LHC & +CEPC  & $\mu^+\mu^-$ & +HL-LHC & +CEPC  \\ 
\hline
$\kappa_W$ & 0.84 & 0.84 & 0.36 & 1.0 & 0.89 &  0.37 \\
$\kappa_Z$ & 0.98 & 0.97 & 0.32 & 1.2 & 1.1  & 0.31 \\
$\kappa_g$ & 1.1  & 1.1  & 0.63 & 1.3 & 1.2  & 0.67 \\
$\kappa_\gamma$ &1.5  & 1.3  & 0.90 & 1.6 & 1.3  & 0.90 \\
$\kappa_{Z\gamma}$ &6.0  & 5.1  & 4.2  & 7.1 & 5.8 &  4.5 \\
$\kappa_c$ & 2.3  & 2.3  & 1.4  & 2.7 & 2.5 &  1.5 \\
$\kappa_t$ & 12.  & 2.9  & 2.6 & 12. &  3.0 &    2.6 \\
$\kappa_b$ & 0.94 & 0.93 & 0.48 & 1.1 & 0.98 &  0.50 \\
$\kappa_\mu$ & 3.5 & 3.0  & 2.5  & 5.6 & 3.7  & 3.1 \\
$\kappa_\tau$ & 1.2 & 1.2  & 0.66 & 1.4 & 1.2  & 0.67 \\
$BR_{BSM}^{95\%}$ & 2.4 & 2.4  & 1.1  & 2.9 & 2.7  & 1.1 \\
\hline
\end{tabular}
\end{center}
\caption{Tabulated results for the 10 TeV $\mu^+\mu^-$ collider off-shell fit in Figure~\ref{fig:KOffshFits}.}\label{KOffshTab10}
\end{table}

\begin{table}[t]
\renewcommand{\arraystretch}{1.5}
\setlength{\arrayrulewidth}{.3mm}
\setlength{\tabcolsep}{1 em}
\begin{center}
\begin{tabular}{|c|c cc|ccc|}
\multicolumn{7}{c}{3 TeV $\mu^+\mu^-$ Collider Off-shell + $BR_{inv}$ Searches Fit Result [\%]} \\ \hline
 & \multicolumn{3}{c|}{With Forward Tagging} & \multicolumn{3}{c|}{Without Forward Tagging} \\ 
 & $\mu^+\mu^-$ & +HL-LHC & +CEPC  & $\mu^+\mu^-$ & +HL-LHC & +CEPC  \\ 
\hline
$\kappa_W$ &  4.8 & 4.7 & 0.57 & 5.0 & 4.8 & 0.59 \\
$\kappa_Z$ &  5.4 & 5.0 & 0.33 & 7.3 & 5.3 & 0.32 \\
$\kappa_g$ &  5.8 & 5.3 & 0.99 & 6.2 & 5.4 & 0.97 \\
$\kappa_\gamma$ &  7.0 & 5.3 & 1.3 & 7.2 & 5.3 & 1.3 \\
$\kappa_{Z\gamma}$ &  25. & 12. & 5.6 & 28. & 12. & 5.6 \\
$\kappa_c$ &  9.5 & 9.7 & 2.0 & 11. & 10. & 1.9 \\
$\kappa_t$ &  40. & 6.3 & 2.6 & 40. & 6.3 & 2.6 \\
$\kappa_b$ &  5.2 & 5.1 & 0.77 & 5.5 & 5.2 & 0.78 \\
$\kappa_\mu$ &  18. & 7.5 & 3.8 & 25. & 7.7 & 3.9 \\
$\kappa_\tau$ &  6.1 & 5.3 & 0.95 & 6.5 & 5.4 & 0.94 \\
$BR_{inv}^{95\%}$ &  13. & 13. & 1.1 & 14. & 13. & 1.1 \\
\hline
\end{tabular}
\end{center}
\caption{Tabulated results for a 3 TeV $\mu^+\mu^-$ collider off-shell + $BR_{inv}$ searches fit.}\label{KBRinv3tab}
\end{table}

\begin{table}[t]
\renewcommand{\arraystretch}{1.5}
\setlength{\arrayrulewidth}{.3mm}
\setlength{\tabcolsep}{1 em}
\begin{center}
\begin{tabular}{|c|c cc|ccc|}
\multicolumn{7}{c}{10 TeV $\mu^+\mu^-$ Collider Off-shell + $BR_{inv}$ Searches Fit Result [\%]} \\ \hline
 & \multicolumn{3}{c|}{With Forward Tagging} & \multicolumn{3}{c|}{Without Forward Tagging} \\ 
 & $\mu^+\mu^-$ & +HL-LHC & +CEPC  & $\mu^+\mu^-$ & +HL-LHC & +CEPC  \\ 
\hline
$\kappa_W$ &  0.71 & 0.70 & 0.34 & 0.80 & 0.74 & 0.36 \\
$\kappa_Z$ &  0.88 & 0.87 & 0.31 & 1.2 & 1.1 & 0.31 \\
$\kappa_g$ &  1.0 & 0.97 & 0.62 & 1.1 & 1.1 & 0.66 \\
$\kappa_\gamma$ & 1.4 & 1.2 & 0.88 & 1.4 & 1.2 & 0.88 \\
$\kappa_{Z\gamma}$ & 6.0 & 5.1 & 4.1 & 6.8 & 5.6 & 4.5 \\
$\kappa_c$ &  2.3 & 2.3 & 1.4 & 2.5 & 2.5 & 1.5 \\
$\kappa_t$ &  12. & 2.8 & 2.6 & 12. & 2.8 & 2.6 \\
$\kappa_b$ &  0.82 & 0.81 & 0.46 & 0.89 & 0.84 & 0.48 \\
$\kappa_\mu$ &  3.4 & 2.8 & 2.5 & 5.4 & 3.6 & 3.2 \\
$\kappa_\tau$ &  1.1 & 1.1 & 0.64 & 1.2 & 1.1 & 0.66 \\
$BR_{inv}^{95\%}$ &  2.1 & 2.1 & 1.0 & 2.4 & 2.2 & 1.0 \\
\hline
\end{tabular}
\end{center}
\caption{Tabulated results for the 10 TeV $\mu^+\mu^-$ collider off-shell + $BR_{inv}$ searches fit in the left chart of Figure~\ref{fig:BRinvFits}.}\label{KBRinv10tab}
\end{table}

\begin{table}[t]
\renewcommand{\arraystretch}{1.5}
\setlength{\arrayrulewidth}{.3mm}
\setlength{\tabcolsep}{1 em}
\begin{center}
\begin{tabular}{|c|c cc|ccc|}
\multicolumn{7}{c}{Off-shell + $ZZ$F $BR_{inv}$ Search Fit Result [\%]} \\ \hline
 & \multicolumn{3}{c|}{3 TeV} & \multicolumn{3}{c|}{10 TeV} \\ 
 & $\mu^+\mu^-$ & +HL-LHC & +CEPC  & $\mu^+\mu^-$ & +HL-LHC & +CEPC  \\ 
\hline
$\kappa_W$ &  0.61 & 0.59 & 0.49  & 0.15 & 0.15 & 0.14 \\
$\kappa_Z$ &  1.4 & 1.0 & 0.26  & 0.38 & 0.37 & 0.14 \\
$\kappa_g$ &  1.9 & 1.3 & 0.88  & 0.49 & 0.48 & 0.42 \\
$\kappa_\gamma$ &  3.4 & 1.3 & 1.2  & 0.88 & 0.69 & 0.70 \\
$\kappa_{Z\gamma}$ &  25. & 8.6 & 5.7  & 5.6 & 4.8 & 4.1 \\
$\kappa_c$ &  6.2 & 6.2 & 1.8  & 1.9 & 1.9 & 1.3 \\
$\kappa_t$ &  38. & 2.6 & 2.6  & 12. & 2.4 & 2.4 \\
$\kappa_b$ &  1.1 & 1.0 & 0.67  & 0.28 & 0.27 & 0.26 \\
$\kappa_\mu$ &  14. & 4.0 & 3.8  & 3.0 & 2.4 & 2.3 \\
$\kappa_\tau$ &  2.4 & 1.3 & 0.83  & 0.64 & 0.59 & 0.42 \\
$BR_{inv}^{95\%}$ &  1.1 & 1.1 & 0.81  & 0.22 & 0.22 & 0.22 \\
\hline
\end{tabular}
\end{center}
\caption{Tabulated fit results for the 3 TeV and 10 TeV $\mu^+\mu^-$ colliders using off-shell and $ZZ$F $BR_{inv}$ search information, the latter corresponding to the right chart in Figure~\ref{fig:BRinvFits}. We have taken $\delta E = 10\%$ for the forward detector energy resolution.}\label{KZZFtab}
\end{table}

\end{document}